\documentclass[aps,pra,twocolumn,superscriptaddress,longbibliography]{revtex4-1} 

\usepackage{amsmath}
\usepackage{amsfonts}
\usepackage{amssymb}
\usepackage{graphicx}
\usepackage{color}
\usepackage{accents}
\usepackage[colorlinks=true, citecolor=cyan]{hyperref}
\usepackage{enumitem}

\newcommand{\Tr}{\mbox{Tr}}

\newcommand{\ket}[1]{|#1\rangle}
\newcommand{\bra}[1]{\langle #1|}

\makeatletter 
\newsavebox{\@brx}
\newcommand{\llangle}[1][]{\savebox{\@brx}{\(\m@th{#1\langle}\)}%
  \mathopen{\copy\@brx\kern-0.5\wd\@brx\usebox{\@brx}}}
\newcommand{\rrangle}[1][]{\savebox{\@brx}{\(\m@th{#1\rangle}\)}%
  \mathclose{\copy\@brx\kern-0.5\wd\@brx\usebox{\@brx}}}
\makeatother

\newlength{\dhatheight} 

\newcommand{\qed}{\nobreak \ifvmode \relax \else
      \ifdim\lastskip<1.5em \hskip-\lastskip
      \hskip1.5em plus0em minus0.5em \fi \nobreak
      \vrule height0.75em width0.5em depth0.25em\fi}

\usepackage{amsmath}
\usepackage{enumitem}
\usepackage{orcidlink}

\begin{document}

\title{Hilbert subspace ergodicity}
\author{Leonard Logari\'c \orcidlink{0000-0001-9486-2047}}
\email[Electronic address: ]{logaricl@tcd.ie}
\affiliation{Department of Physics, Trinity College Dublin, Dublin 2, Ireland}
\affiliation{Trinity Quantum Alliance, Unit 16, Trinity Technology and Enterprise Centre, Pearse Street, Dublin 2, D02 YN67, Ireland}
\author{John Goold \orcidlink{0000-0001-6702-1736}}
\email[Electronic address: ]{gooldj@tcd.ie}
\affiliation{Department of Physics, Trinity College Dublin, Dublin 2, Ireland}
\affiliation{Trinity Quantum Alliance, Unit 16, Trinity Technology and Enterprise Centre, Pearse Street, Dublin 2, D02 YN67, Ireland}
\affiliation{Algorithmiq Limited, Kanavakatu 3C 00160 Helsinki, Finland}
\author{Shane Dooley \orcidlink{0000-0002-2856-8840}}
\email[Electronic address: ]{dooleysh@gmail.com}
\affiliation{Dublin Institute for Advanced Studies, School of Theoretical Physics,
10 Burlington road, Dublin, D04 C932, Ireland}
\affiliation{Trinity Quantum Alliance, Unit 16, Trinity Technology and Enterprise Centre, Pearse Street, Dublin 2, D02 YN67, Ireland}


\date{\today}

\begin{abstract}
Ergodicity has been one of the fundamental concepts underpinning our understanding of thermalization in isolated systems since the first developments in classical statistical mechanics. Recently, a similar notion has been introduced for quantum systems, termed complete Hilbert space ergodicity (CHSE), in which the evolving quantum state explores all of the available Hilbert space. This contrasts with the eigenstate thermalization hypothesis (ETH), in which thermalization is formulated via the properties of matrix elements of local operators in the energy eigenbasis. In this work we explore how ETH-violation mechanisms, including quantum many-body scars and Hilbert space fragmentation can affect complete Hilbert space ergodicity. We find that the presence of these mechanisms leads to CHSE in decoupled subspaces, a phenomenon we call Hilbert subspace ergodicity, and which represents a protocol for constructing $t$-designs in subspaces.
\end{abstract}

\maketitle

\section{Introduction}\label{sec:introduction}

One of the key developments in statistical physics has been the introduction of ergodicity. A classical system is deemed to be ergodic if all of the available points in phase space are explored with equal probability over long timescales \cite{Bol-1896}. Such a notion however, \emph{appears} to be inconsistent with dynamics of closed quantum systems,  due to the constraints imposed by their unitary evolution. This has spurred significant research interest into how thermalization can be understood in isolated quantum systems. The most widely accepted framework in this area is the eigenstate thermalization hypothesis (ETH) \cite{Deu-91a, Sre-94a, Rig-08a, Deu-18a}. The ETH states that the apparent thermalization of quantum systems occurs due to the fact that eigenstates of large, generic many body systems appear indistinguishable from thermal states at the level of local observables. The formulation of ETH also led to the exploration of different forms of ETH-violation, which can be subdivided into \emph{strong} and \emph{weak ETH-violation}. The former include systems which exhibit integrability \cite{Cal-16}, many-body localisation (MBL) \cite{Nan-15a, Aba-19}, or strong Hilbert space fragmentation (HSF) \cite{Sal-20, Mou-22a, Mou-22c, Han-24}, while the latter refer to systems with quantum many-body scars (QMBS) or weak HSF \cite{Tur-18a, Pap-22a, Cha-23a, Ser-21a, Des-22a, Doo-21a, Doo-23a}. 
Both phenomena have also been investigated in digital quantum circuits \cite{Han-24, Log-24a, And-24}, as these have been rising to prominence both due to their usage to probe interesting physical phenomena \cite{Vov-21, Sta-22, Sum-23, Kee-23}, as well as fascinating systems on their own \cite{Ber-18, Ber-19, Sum-24}. 

In contrast to ETH, recently Pilatewski \emph{et al}. \cite{Pil-23a} introduced the concept of complete Hilbert space ergodicity (CHSE). This notion of ergodicity is defined at a dynamical level, and refers to the phenomenon by which any initial state will explore all other states within the Hilbert space with ``equal probability'' in the long time limit. This opens up several research questions. What is the effect of ETH-violating mechanisms on the dynamics of the systems which would otherwise exhibit CHSE? What is the effect of symmetries? Can the notion of CHSE be generalized to these cases, where the many-body systems contain conserved quantities?

\begin{figure}
    \centering
    \includegraphics[width=\linewidth]{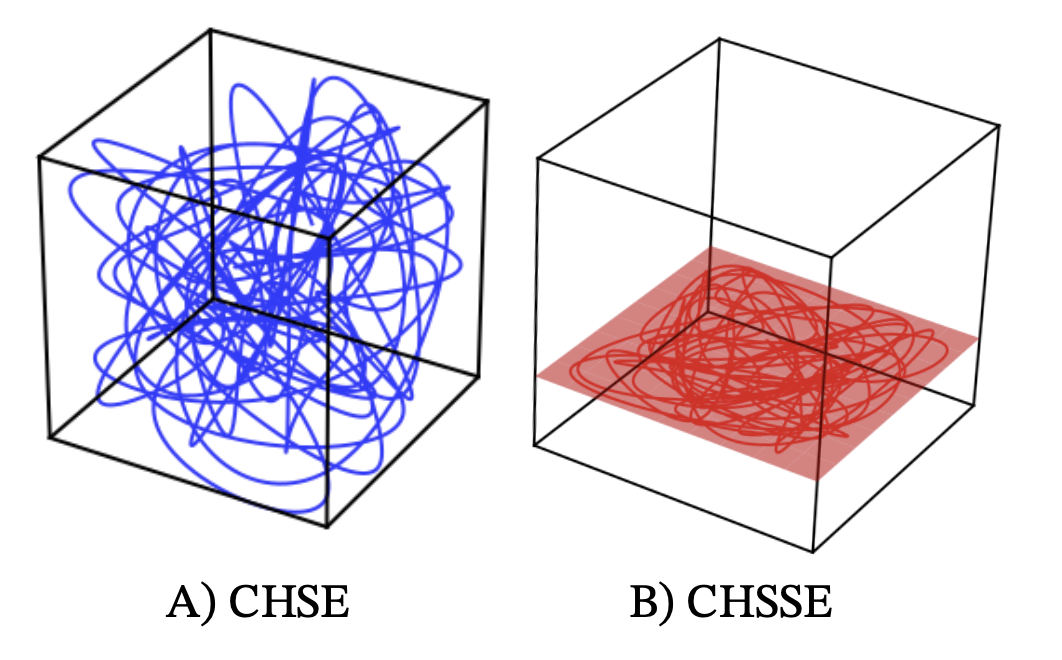}
    \caption{Schematic representation for complete Hilbert space
    ergodicity (A) vs. complete Hilbert space ergodicity in a subspace (B). In the former, the system explores all of the available $D$-dimensional Hilbert space, represented by a three-dimensional box. In the latter however, the system can only explore a $B$-dimensional subspace of the full Hilbert space, represented by a two-dimensional horizontal slice. In the long time limit, the system represented in panel (A) will explore all regions within the box with equal probability, whereas the system in panel (B) will explore just the shaded surface.}
    \label{fig:chse_v_chsse_comparison}
\end{figure}

We aim to answer these questions by analyzing the dynamical properties of simple circuit models, composed of $2$-local quantum gates, and showing the consequences of the presence of scars, fragmentation and symmetries. It is accepted that in scarred and fragmented models certain initial states do not thermalise, and it immediately follows that such models cannot exhibit CHSE. However, the implications on the ``thermalizing'' subspaces are not self-evident. Our main finding is that, in the presence of nonlocal conserved quantities, generic aperiodically driven systems can display CHSE within different subspaces, which we call Hilbert Subspace Ergodicity, schematically depicted in Fig. \ref{fig:chse_v_chsse_comparison}.

This paper is organized as follows. In Sec. \ref{sec:background} we provide a brief overview of the ETH and two ETH-violation mechanisms, QMBS and Hilbert space fragmentation. We also outline the notion of complete Hilbert space ergodicity. Section \ref{sec:chsse} describes in detail the introduced notion of complete Hilbert subspace ergodicity. In Sec. \ref{sec:model} we present the model used for all our numerical evidence, and in Sec. \ref{sec:bw_chse} we show that it exhibits CHSE in the absence of any conserved quantities. We demonstrate the effects of scars in an otherwise CHSE system in Sec. \ref{sec:scars}, and the effects of fragmentation in Sec. \ref{sec:fragmentation}. For completeness, we also cover the effect of symmetries in Sec. \ref{sec:symmetries}. Finally, in Sec. \ref{sec:conclusion} we give a brief overview of results and their implications, as well as discuss potential avenues for future research.

\section{Background}\label{sec:background}

\subsection{Complete Hilbert space ergodicity}

Suppose that a state $\ket{\psi(0)}$ in a $D$-dimensional Hilbert space undergoes unitary dynamics, $\ket{\psi(t)} = \hat{\mathbb{U}}(t)\ket{\psi(0)}$. For concreteness, we assume discrete-time dynamics where $\ket{\psi(t)} = \hat{\mathbb{U}}_t\ket{\psi(t-1)}$ propagates the system forward one time-step from $t-1$ to $t$ and $\hat{\mathbb{U}}(t) = \hat{\mathbb{U}}_t \hat{\mathbb{U}}_{t-1} \hdots \hat{\mathbb{U}}_1$. In analogy with the classical notion of ergodicity, CHSE means that given sufficient evolution time the quantum dynamics explores all of its available Hilbert space with equal likelihood. To be precise about the idea of ``equal likelihood,'' one must introduce a measure on the Hilbert space. The natural choice is the unique measure ${\rm d}\phi$ that is invariant with respect to all unitary transformations, called the Haar measure. More formally then, CHSE means that \cite{Pil-23a}: \begin{equation} \lim_{T\to\infty} \frac{1}{T} \int_0^T dt f (\ket{\psi(t)}\bra{\psi(t)}) = \int {\rm d}\phi f (\ket{\phi}\bra{\phi}) , \label{eq:CHSE} \end{equation} for any initial state $\ket{\psi(0)}$ and for any integrable function $f$. In other words, if the system is CHSE then the long-time average of any observable quantity is indistinguishable from its Haar average.

In practice one cannot test Eq.~\eqref{eq:CHSE} for all integrable functions $f$ and it is more convenient to restrict to polynomial functions of degree $k$, which can always be computed using a limited set of moments. Following Ref.~\cite{Pil-23a}, we define the $k$'th moment of the ensemble of states in the timeseries $\{ \ket{\psi(t)} \}_{t=0}^{T-1}$ as: \begin{equation} \hat{\rho}_T^{(k)} = \frac{1}{T}\sum_{t=0}^{T-1} (\ket{\psi(t)}\bra{\psi(t)})^{\otimes k} , \label{eq:k_moment_time} \end{equation} and the $k$'th moment of the Haar ensemble of states as: \begin{equation} \hat{\rho}_{\rm Haar}^{(k)} = \int {\rm d}\phi (\ket{\phi}\bra{\phi})^{\otimes k}. \label{eq:k_moment_haar} \end{equation} We note that the $k$'th moment of the Haar ensemble can be computed exactly and has a closed form expression~\cite{Har-13a}:

\begin{equation}
    \hat{\rho}^{(k)}_{\text{Haar}} = \frac{\sum_{\pi \in \mathcal{S}_k} \sum _{\alpha_1, ..., \alpha_k} | \alpha_1, \alpha_2, \hdots  \alpha_k \rangle \langle \alpha_{\pi(1)}, \hdots  \alpha_{\pi(k)}|}{D (D + 1) ... (D + k - 1)}
    \label{eq: haar_moments}
\end{equation}
where $\mathcal{S}_k$ is the permutation group of order $k$ and $\{ \ket{\alpha_i} \}_{i=1}^D$ is an orthonormal basis for the Hilbert space of dimension $D$. 

We denote the Hilbert-Schmidt distance between $\hat{\rho}_T^{(k)}$ and $\hat{\rho}^{(k)}_{\text{Haar}}$ as: 
\begin{equation} 
\Delta_T^{(k)} = D_{\text{HS}}(\hat{\rho}^{(k)}_{\text{Haar}}, \hat{\rho}^{(k)}_T) = \| \hat{\rho}^{(k)}_{\text{Haar}} - \hat{\rho}^{(k)}_T  \|_2^2, 
\label{eq: Delta_Tk} 
\end{equation} 
where $\| \hat{X} \|_2 = \sqrt{{\rm Tr}(\hat{X}^\dagger\hat{X})}$ is the Hilbert-Schmidt matrix norm \footnote{We note that Ref. \cite{Pil-23a} used the trace distance instead of the Hilbert-Schmidt distance in Eq.~\eqref{eq: Delta_Tk}. The trace distance has the the advantage of a clear operational meaning in quantum state discrimination. However, to calculate it requires matrix diagonalisation while the Hilbert-Schmidt distance does not, and is easier to calculate numerically.}. Note that when originally introduced in Ref. \cite{Pil-23a}, the CHSE was probed using the trace distance as a metric in Eq.~\eqref{eq: Delta_Tk}. However, since the Hilbert-Schmidt distance can be used to upper bound the trace distance \cite{Col-19a} and is significantly computationally cheaper to compute we will use the former. Now, the system is CHSE if $\Delta_\infty^{(k)} = 0$ for all $k$ and for all initial states $\ket{\psi(0)}$. 

Furthermore, in the case of a time-independent Hamiltonian, one can show that for $k \geq 2$ the trace distance to the Haar moments cannot be lower than \cite{Pil-23a}:
\begin{equation}
    B (D) \equiv (D + 1)^{-1} - (2D (D+1)) ^{-\frac{1}{2}}, 
\end{equation}
which one can adapt for the Hilbert-Schmidt distance. Since $D_{\text{Tr}}(\rho, \sigma)^2 \leq \frac{D}{4} D_{\text{HS}} (\rho, \sigma)$, we have that the lower bound for the Hilbert-Schmidt distance is: \begin{equation} \Delta_T^{(k)} \geq \frac{4B^2}{D}, \label{eq:Delta_LB} \end{equation} for time-independent Hamiltonian dynamics.

Another quantity to probe CHSE is the \emph{ensemble entropy}, recently introduced in Ref.~\cite{Mar-24}. The key idea is to directly compare the states sampled by time evolution to states sampled from the Haar distribution. An ensemble is a collection of pure quantum states with associated probabilities:
\begin{equation}
    \label{eq:ensemble}
    \mathcal{E} = \{p_j, \ket{\Psi _j} \},
\end{equation}
where $j$ can be a discrete value (corresponding to, for example, different measurement outcomes) or a continuous one (such as time). In the case of discretised probability distributions we will use index $j$ to refer to the different states, but in case of continuous variables we will use $\Psi$ itself as the variable and write the ensemble as: $\mathcal{E} = \{P(\Psi), \ket{\Psi } \}$

We will denote the Haar ensemble as $\mathcal{E}_{\text{Haar}} = \{P^H(\Psi), \ket{\Psi} \}$, which will be the reference ensemble to compute the ensemble entropy. In the case of a discretized version of the Haar ensemble, we will use the notation $\mathcal{E}_{\text{Haar}} = \{p_j^H, \ket{\Psi _j} \}$. The ensemble generated by some aperiodic time evolution, also called the temporal ensemble, will be denoted as $\mathcal{E}_T = \{P^T (\Psi), \ket{\Psi} \}$ for the continuous version, and as $\mathcal{E}_T = \{p_j^T, \ket{\Psi _j ^T} \}$ for discretized one. The ensemble entropy for this generated temporal ensemble, is then defined as the negative of the Kullback-Leibler divergence: 

\begin{equation} \label{eq: ens_ent}
    \text{Ent}(\mathcal{E}_T | \mathcal{E}_{\text{Haar}} )  = - \int d\Psi P^T(\Psi) {\log}_2 \left( \frac{P^T(\Psi)}{P^H (\Psi )}\right).
\end{equation}
This quantity directly compares the two ensembles, and is exactly $0$ if and only if they are identical. If that holds, it also immediately follows that the Hilbert-Schmidt distance between the moments constructed from these two ensembles, $\hat{\rho}_{\text{Haar}}^{(k)}$ and $\hat{\rho}_T^{(k)}$, will also vanish for any $k$.

In practice, when numerically evaluating this measure we are using a finite set of states generated by the time evolution, $\{ \ket{\psi (t)}\}_{t = 0}^{T-1}$, which demands a discretisation of Eq.~\eqref{eq: ens_ent}, replacing the integral with a discrete summation over a finite set of states and associated probabilities, which we term as discretized ensemble entropy (DEE): 

\begin{equation} \label{eq: DEE}
    \text{DEE}(\mathcal{E}_T | \mathcal{E}_{\text{Haar}}) = - \sum_{j = 1}^M p_j^T \log _2 (\frac{p_j^T}{p^H_j}). 
\end{equation}
Further details about how we obtained the discretization of the Haar ensemble and the numerical calculation can be found in Appendix \ref{app: ensemble entropy}.

\subsection{Quantum many-body Scars and Hilbert space fragmentation}

One of the most important concepts to understand thermalization of closed quantum systems is the ETH \cite{Deu-91a, Sre-94a, Rig-08a, Deu-18a}. This phenomenon has been extensively studied and verified both numerically \cite{Bre-20a, Ste-13} and experimentally \cite{Tro-12, Clo-16, Kau-16, Tan-18}.

However, in a seminal experiment performed on a 51-qubit array of Rydberg atoms, it was found that thermalization fails when the system is initialised in specific initial states \cite{Ber-17}. The reason for the failure to thermalise is that the special initial states have a significant overlap with a small subset of ETH-violating eigenstates called quantum many body scars (QMBS) \cite{Tur-18a}. Since the first experiment, several more experiments have found evidence of QMBS in different systems \cite{Zha-22, Su-23}. Also, a large amount of theoretical work has uncovered QMBS in various models, and laid the foundations for several scarring mechanisms \cite{Mou-22c}. Two of the most prominent approaches to encapsulate a wide range of scarred models are the projector-embedding \cite{Shi-17} and the spectrum generating algebra (SGA) formalisms \cite{Mou-20a}.

Another mechanism of ETH-violation is the so-called Hilbert space fragmentation (HSF), first explicitly identified in fracton models \cite{Pai-19, Sal-20, Khe-20}. In models with HSF, the Hilbert space ``shatters'' into exponentially many fragments, with respect to system size. Based on the scaling of the dimensions of these fragments, one can further distinguish between \emph{strong} and \emph{weak} fragmentation, which lead to \emph{strong} and \emph{weak} ETH-violation, respectively. Since weak HSF is very similar to QMBS in many aspects, it raised questions about the fundamental differences between these two phenomena. Ultimately, both of these mechanisms came to be understood within the framework of commutant algebras \cite{Mou-22a, Mou-22b}.

\begin{figure}
    \centering
    \includegraphics[width=\linewidth]{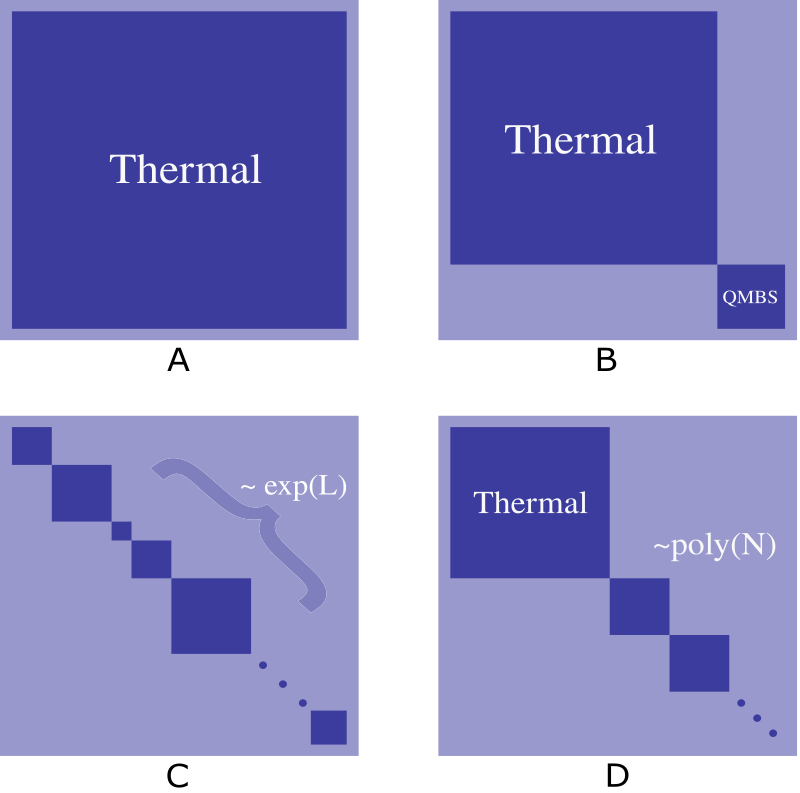}
    \caption{Schematic representation of the dynamically decoupled Krylov subspaces for chaotic systems (A), systems with QMBS (B), Hilbert space fragmentation (C), and conventional symmetries (D). (A) In the case of chaotic systems, all states are thermal. (B) For scarred systems, the majority of the states are thermal, while there is a vanishingly small number of QMBS states, with respect to system size. (C) For systems with Hilbert space fragmentation, the number of decoupled subspaces scales exponentially with system size. The fundamental difference between systems with fragmentation and symmetries (D), is that in the former the number of decoupled subspaces is exponential in system size, while in the latter it is constant or at most polynomial. }
    \label{fig:subspace_decoupling}
\end{figure}

Common to both QMBS and HSF is the property that, starting from simple-to-prepare initial states, the dynamics can fail to lead to thermalization. This is because some simple initial states can be prepared that exist entirely in a disconnected, nonthermal subspace: in the case of models with QMBS, the scar states are effectively decoupled from the rest of the spectrum, and in models with HSF, there are exponentially many of these decoupled subspaces. This is schematically depicted in Fig.~\ref{fig:subspace_decoupling}. Therefore both the Hamiltonian $\hat{H}$ and its corresponding unitary $\hat{U}$ will have a block diagonal structure. We will refer to these dynamically decoupled subspaces, represented by the different ``blocks'', as Krylov subspaces $\mathcal{K}_\alpha$. They are defined through the time evolution from different simple-to-prepare initial states $\ket{\psi_\alpha}$ (e.g., computational basis states) \cite{Mou-22a}:
\begin{equation}
    \mathcal{K}_{\alpha} \equiv \text{span}_t \{\hat{U}^{t} \ket{\psi _{\alpha}}\}
\end{equation}
Note that the time evolution of different initial states may generate the same subspace. Therefore the full Hilbert space can be expressed as:
\begin{equation}
    \mathcal{H} = \bigoplus_{\alpha  = 1}^{\mathcal{N}_{\mathcal{K}}} \mathcal{K}_{\alpha},
\end{equation}
where $\mathcal{N}_{\mathcal{K}}$ is the number of unique Krylov subspaces.

The number of distinct Krylov subspaces $\mathcal{N}_{\mathcal{K}}$ distinguishes between QMBS and HSF, while the size of the largest subspace, $D_{\text{max}}$, distinguishes between \emph{strong} and \emph{weak} fragmentation. If we have a constant, or polynomial number of Krylov subspaces, then the system contains QMBS. If, however, the number of Krylov subspaces is exponential in system size, then we can conclude the system exhibits HSF. Furthermore, if the dimension of the largest subspace, $D_{\text{max}}$, is a nonvanishing fraction of the total Hilbert space dimension, $\lim_{N \rightarrow \infty} (D_{max}/\text{dim}(\mathcal{H}) )> 0$, then it is an instance of weak fragmentation, whereas if this fraction is vanishing, $\lim_{N \rightarrow \infty} (D_{max}/\text{dim}(\mathcal{H})) = 0$, then it is an example of strong fragmentation \cite{Sal-20}. Signatures of strong and weak violation of the ETH can also be observed in the growth of the bipartition entanglement entropy and the decay of the infinite temperature autocorrelator, which are discussed in more detail in Appendices \ref{app:autocorrelators} and \ref{app:bipartite_entanglement_entropy}. 

\section{Complete Hilbert Subspace Ergodicity (CHSSE)} \label{sec:chsse}

The original notion of CHSE was introduced for generic aperiodic drives without any conserved quantities. However, in the presence of (local or nonlocal) conserved quantities the Hilbert space may decouple into several Krylov subspaces $\mathcal{K}_{\alpha}$. Because of this decoupling, it will be impossible to exhibit CHSE in the full Hilbert space. However, the system may still ``explore'' the subspaces with equal probability, which we term complete Hilbert subspace ergodicity (CHSSE).

Consider some finite, $D$-dimensional Hilbert space. The dynamics of this system is governed either by some time dependent Hamiltonian $\hat{H}(t)$, with $t \in \mathbb{R}$, or a sequence of unitary operations $\mathbb{\hat{U}}_t$, with $t \in \mathbb{N}$. The Hamiltonian or sequence of unitaries splits into $\mathcal{N}_{\mathcal{K}}$ dynamically disconnected Krylov subspaces $\mathcal{K}_{\alpha}$, but is otherwise completely generic. A system will display CHSSE provided that, starting in any initial state in some Krylov subspace, $\ket{\psi (t = 0)} \in \mathcal{K}_{\alpha}$, the $k$-th moment of the temporal ensemble, $\hat{\rho}_T^{(k)}$, will converge to the $k$-th moment of the Haar ensemble in the $\mathcal{K}_{\alpha}$ subspace, defined as:
\begin{eqnarray}
    \hat{\rho}_{\mathcal{K}_{\alpha}\text{-Haar}}^{(k)} &=& \int _{\ket{\phi} \in \mathcal{K}_{\alpha}} {\rm d}\phi (|\phi \rangle \langle \phi | )^{\otimes k} \notag
    \\ &=&  \frac{\sum\limits_{\pi \in \mathcal{S}_k} \sum \limits_{ \{\beta_i \}} | \beta_{1}, ...,  \beta_{k}\rangle \! \langle \beta_{\pi(1)}, ...,  \beta_{\pi(k)}|}{D_{\alpha} (D_{\alpha} + 1) ... (D_{\alpha} + k - 1)},
    \label{eq: subspace-haar}
\end{eqnarray}
where $\pi$ are permutation operators belonging to the permutation group $\mathcal{S}_k$, $D_{\alpha} = \text{dim}(\mathcal{K}_{\alpha})$ and $\beta_i$ label different basis states in the Krylov subspace. Note that in this case ${\rm d}\phi$ is the unique unitary-invariant measure in the $\mathcal{K}_{\alpha}$ subspace. 
For time-independent Hamiltonian dynamics, due to the constraint of energy conservation (and the commutant algebra set), we again have a lower bound on the trace distance (and consequently also Hilbert-Schmidt distance) between the $k$'th moments of the temporal ensemble and the Haar ensemble in the decoupled subspaces. By the same argument as in Ref. \cite{Pil-23a}, the bound on the trace distance between the $k \geq 2$ moments of the temporal ensemble generated by some initial state $\ket{\psi _{\alpha}} \in \mathcal{K}_{\alpha}$ and  $\hat{\rho}^{(k)}_{\mathcal{K}_{\alpha}-\text{Haar}}$ will be $B(D_{\alpha})$. 

\section{Setup}\label{sec:model}

\begin{figure}
    \centering
    \includegraphics[width=0.9\columnwidth]{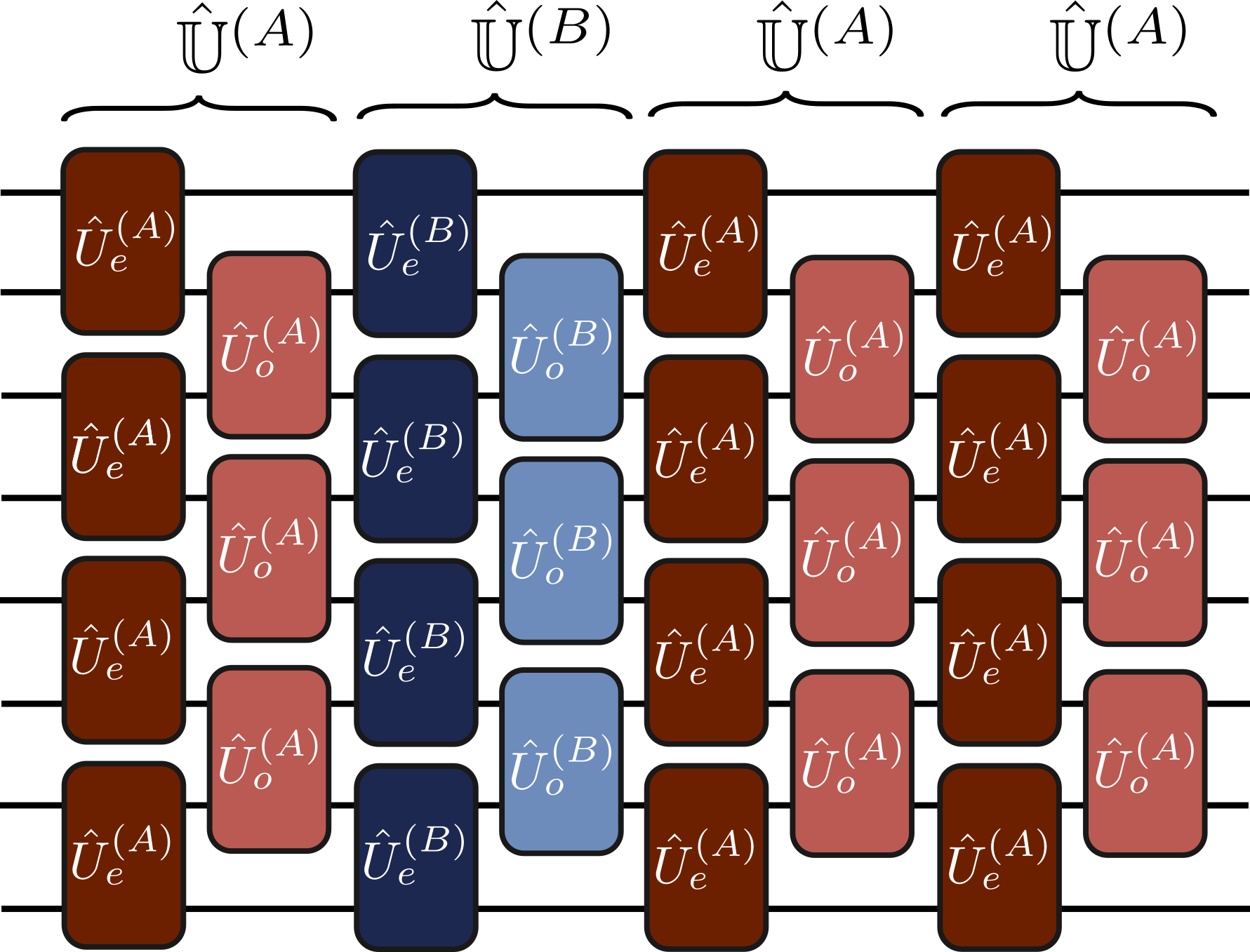}
    \caption{The first four timesteps of our brickwork circuit for $N=8$. The aperiodic sequence of unitaries $\hat{\mathbb{U}}^{(A)} \hat{\mathbb{U}}^{(A)} \hat{\mathbb{U}}^{(B)} \hat{\mathbb{U}}^{(A)}$ is determined through the Fibonacci word.}
    \label{fig:brickwork}
\end{figure}

To investigate the potential effects of the presence of scars and fragmentation on CHSE, we consider a chain of $N$ qudits, labeled by $i \in \{0, 1, 2, ..., N-1 \}$, each with local Hilbert space dimension $d$. The chain of qudits evolves dynamically by a ``brickwork'' quantum circuit, made up of layers of two-qubit nearest-neighbor unitaries. We choose our brickwork circuit so that each timestep is completely specified by the two ``bricks'' $\hat{U}_e = \exp (i\hat{H}_e)$ and $\hat{U}_o = \exp (i\hat{H}_o)$, where $\hat{U}_e$ makes up the even layer $\hat{\mathbb{U}}_e = \hat{U}_e \otimes \hat{U}_e \otimes \hdots $ of the brickwork, and $\hat{U}_o$ makes up the odd layer $\hat{\mathbb{U}}_o = \hat{\mathbb{I}} \otimes \hat{U}_o \otimes \hat{U}_o \otimes \hdots$ (see Fig. \ref{fig:brickwork} for an illustration). The system is then propagated forward one step in time by a unitary operator of the form: 
\begin{equation} \label{eq:U_brick} \hat{\mathbb{U}} = \hat{\mathbb{U}}_o \hat{\mathbb{U}}_e . \end{equation}

However, to generate CHSE, our circuit cannot be periodic in time. So, we introduce two different unitary propagators: $\hat{\mathbb{U}}^{(A)}$ [specified by the local unitary bricks $\hat{U}_e^{(A)}$, $\hat{U}_o^{(A)}$] and $\hat{\mathbb{U}}^{(B)}$ [specified by the bricks $\hat{U}_e^{(B)}$, $\hat{U}_o^{(B)}$]. We can apply these two different unitaries in a sequence that is deterministic but also completely aperiodic, through the Fibonacci word $W_j$, which is defined recursively \cite{Pil-23a}. We set two initial words $W_0 = 1$, $W_1 = 0$, and then construct longer words using the rule $W_{j + 1} = W_j W _{j - 1}$. This rule can be repeated infinitely many times to give $W_{\infty}$. Using $W_{\infty}$ we then implement the discrete time evolution as a sequence of unitaries $\hat{\mathbb{U}}^{(A)}$ or $\hat{\mathbb{U}}^{(B)}$: at a given time $t$ we apply $\hat{\mathbb{U}}^{(A)}$ if the $t$-th symbol in $W_{\infty}$ is $0$, and $\hat{\mathbb{U}}^{(B)}$ if it is $1$. The resulting unitary then has the form: \begin{equation} \hat{\mathbb{U}}(t) = \hdots \hat{\mathbb{U}}^{(B)} \hat{\mathbb{U}}^{(A)} \hat{\mathbb{U}}^{(B)} \hat{\mathbb{U}}^{(A)} \hat{\mathbb{U}}^{(A)} \hat{\mathbb{U}}^{(B)} \hat{\mathbb{U}}^{(A)}           . \end{equation} The brickwork circuit with this aperiodic sequence of gates will be used throughout this work. For the system to exhibit scars or fragmentation, we impose specific constraints on the local bricks $\hat{U}_e^{(A)}$, $\hat{U}_o^{(A)}$, $\hat{U}_e^{(B)}$, and $\hat{U}_o^{(B)}$.

In Ref. \cite{Pil-23a} it was shown that for generic $\hat{\mathbb{U}}^{(A)}$ and $\hat{\mathbb{U}}^{(B)}$, this Fibonacci driving leads to CHSE. Next, we will verify that the model illustrated in Fig. \ref{fig:brickwork} also exhibits CHSE for generic $2$-local unitaries. This complements a result of Ref. [29], which shows CHSE for an aperiodically driven spin chain model with local connectivity.

\begin{figure}
    \centering
    \includegraphics[width=\linewidth]{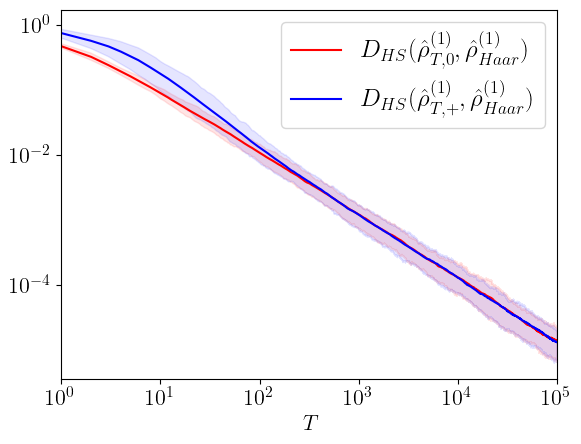}
    \includegraphics[width=\linewidth]{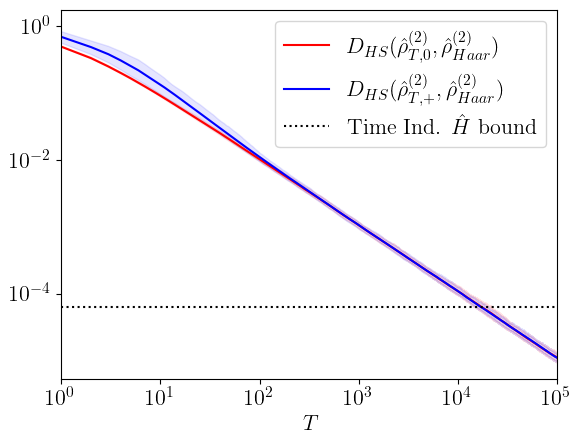}
    \caption{Hilbert-Schmidt distance between the Haar moments and the temporal ensemble moments for $k = 1$ (top) and $k = 2$ (bottom). The results are averaged over 100 different instances of $\{\hat{\mathbb{U}}^{(A)}, \hat{\mathbb{U}}^{(B)} \}$, with the lines corresponding to the average distances, and the shaded areas represent the 10th-90th percentile range. The red plot corresponds to the initial state $\ket{\psi (0)} = \ket{0}^{\otimes N}$, and the blue one to $\ket{\psi (0)} = \ket{+}^{\otimes N}$. The dotted line in the lower plot for $k = 2$ represents the lower bound for time independent Hamiltonians, and is surpassed for both initial states.}
    \label{fig:bw_k1}
\end{figure}

\section{CHSE in a Brickwork Model with Generic Local Unitaries} \label{sec:bw_chse}

First, we show that our circuit model (as illustrated in Fig. \ref{fig:brickwork}) is CHSE for generic instances of the local unitary bricks. We provide a detailed argument in \ref{app: universality}, and in this section we provide numerical evidence. To show this, we consider a simple instance of the model: we use $N = 4$ qubits ($d = 2$) with open boundary conditions. The dynamics are defined by the four local unitary bricks $\{ \hat{U}_e^{(A)}, \hat{U}_{o}^{(A)}, \hat{U}_{e}^{(B)}, \hat{U}_{o}^{(B)} \}$. Each of the local unitaries is created using a random Hermitian generator $\hat{H}_{o/e}^{(A/B)}$, where the real and imaginary parts of the matrix elements are sampled uniformly from the interval $[0, 1)$, and then constructing the unitary as $\hat{U}_{o/e}^{(A/B)} = \exp (-i \hat{H}_{o/e}^{(A/B)})$. We sample over 100 different instances of the circuit, using different sets of local unitaries, $\{ \hat{U}_e^{(A)}, \hat{U}_{o}^{(A)}, \hat{U}_{e}^{(B)}, \hat{U}_{o}^{(B)} \}$, for each of them. 

We then compute the Hilbert-Schmidt distances to the Haar moments for $k = 1, 2$, using two different initial states: $\ket{\psi (0)} = \ket{0}^{\otimes N}$ and $\ket{\psi (0)} = \ket{+}^{\otimes N}$. At each time-step we compute the first and second moments of the temporal ensemble, $\hat{\rho}_{T, 0/+}^{1/2}$, where the second subscript denotes the corresponding initial state.

\begin{figure}
    \centering
    \includegraphics[width=\linewidth]{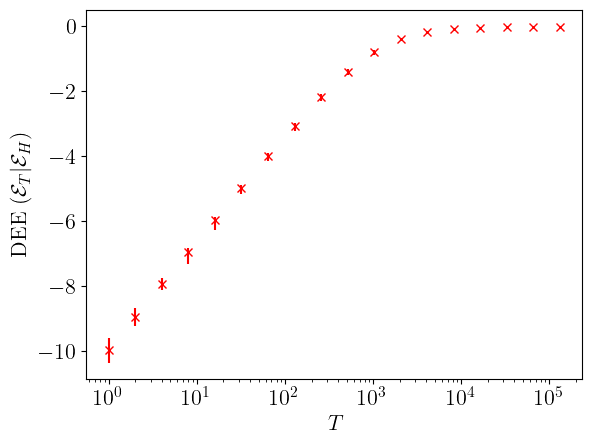}
    \caption{ Discretized ensemble entropy for the dynamics generated by a generic brickwork circuit. Note that we have computed it for the initial state $\ket{0}^{\otimes N}$, and for a single instance of $\{ \hat{\mathbb{U}}^{(A)}, \hat{\mathbb{U}}^{(B)}\}$. We compute the DEE using 100 different instances of sets of reference states $\{\ket{\psi_j} \}$, with $M \approx 1000$ states in each set. The error bars represent the spread between minimum and maximum values, and the crosses display the mean. The discretized ensemble entropy is converging to $0$ as the number of states generated by the dynamics, $T$, is increased.}
    \label{fig:gbw_ensemble_entropy}
\end{figure}

The results are presented in Fig. \ref{fig:bw_k1}. As a first check, for $k = 1$ we observe that the ensembles generated from both the $\ket{\psi (0)} = \ket{0}^{\otimes N}$ and the $\ket{\psi (0)} = \ket{+}^{\otimes N}$ initial states converge to the first moment of the Haar ensemble. For both initial states we also observe a power law decay of the distances, $\Delta_T^k \sim T^{-1/2}$, which was already observed in Ref. \cite{Pil-23a}. These results for $k=1$ are not particularly surprising, since any Floquet driving without energy conservation will ``heat up'' any initial state to the infinite temperature state $\hat{\rho} = \frac{1}{D} \hat{\mathbb{I}}$, which is equal to the first Haar moment: $\hat{\rho}_{\text{Haar}}^{(1)} = \int d\phi \ket{\phi} \bra{\phi} = \frac{1}{D} \hat{\mathbb{I}}$. To numerically investigate CHSE we should check the distance between $\hat{\rho}_{T}^{(k)}$ and $\hat{\rho}_{\textbf{}{Haar}}^{(k)}$ for higher moments $k>1$.

Comparing the $k=2$ moments of the temporal and Haar ensembles in Fig. \ref{fig:bw_k1}, we again see strong evidence for CHSE. Both initial states give very similar long time behavior, overlapping almost perfectly for times $T > 10^2$. Interestingly, they display even smaller variation between the different instances of the model than for the $k=1$ moment, with all of the results in the 10th-90th percentile range strongly concentrated around the mean values. It is clear that the distances to the Haar moment go below the bound for time-independent Hamiltonians (Eq.~\eqref{eq:Delta_LB}) \cite{Pil-23a}. 

For conclusive numerical evidence that these models do exhibit CHSE one would have to go to higher moments $k>2$ and longer times $T$. However, our numerical results for $k=2$, together with the argument in Appendix \ref{app: universality}, demonstrate CHSE for generic 2-local brickwork circuit models. To corroborate this further beyond the $k=2$ moment, we also compute the DEE for the initial state $\ket{0}^{\otimes N}$, shown in Fig. \ref{fig:gbw_ensemble_entropy}. The DEE converges to $0$ with increasing temporal set size $T$, as expected for CHSE.

\section{CHSSE through Quantum Many-Body Scars}\label{sec:scars}

\begin{figure*}
    \centering
    \includegraphics[width=0.44\linewidth]{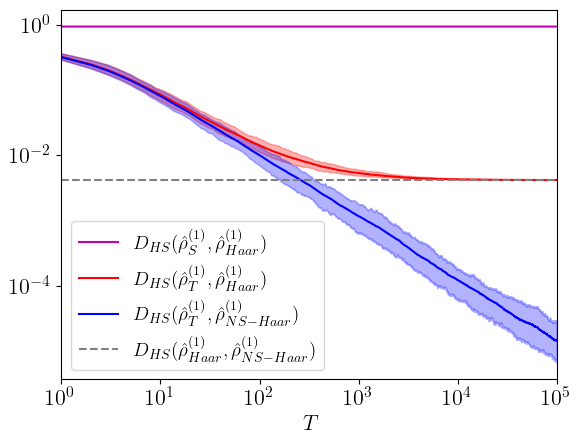}
    \includegraphics[width=0.44\linewidth]{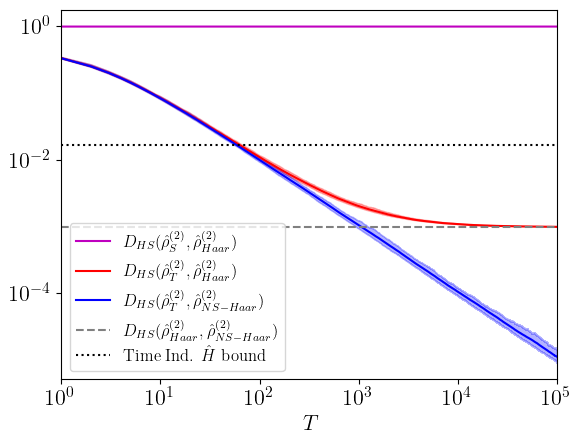}
    \caption{Effect of embedding a single QMBS into a brickwork model. The results are obtained by using 100 different instances of $\{\hat{\mathbb{U}}^{(A)}, \hat{\mathbb{U}}^{(B)} \}$. Left: Distance between the first moments of the temporal ensembles and the Haar ensemble. Right: Distance between the second moments of the temporal ensembles and the Haar ensemble. The violet line represents the results for the initial scar state $\ket{S_1} = \ket{0}^{\otimes N}$. Since it is invariant under the applied unitaries, the generated moments ($\hat{\rho}_{S}^{(k = 1, 2)}$) are constant, and so is their distance to the respective Haar moments. The red line represents the average distance between the moments of the temporal ensemble ($\hat{\rho}_{T}^{(k = 1,2)}$) and Haar ensemble ($\hat{\rho}_{\text{Haar}}^{(k = 1,2)}$) for the $\ket{1}^{\otimes N}$ initial state, and the shaded region represents the 10th-90th percentile region. We observe that for both moments, this distance is bounded from below, which is expected due to the separation of the scar and nonscar subspaces. To demonstrate this, we construct the first and second moments of the Haar ensemble constrained to the nonscar subspace, ($\hat{\rho}_{\text{NS-Haar}}^{(k = 1,2)}$), by using Eq.~\eqref{eq: subspace-haar} with subspace $\mathcal{K}_{\text{nonscar}} = \mathcal{H} \; \setminus \; \mathcal{H}_S $. The dashed lines represent the distances between the full Hilbert space ($\hat{\rho}_{\text{Haar}}^{(k)}$) and nonscar subspace ensembles ($\hat{\rho}_{\text{NS-Haar}}^{(k)}$) for the first and second moments ($k = 1, \; 2$), and indicate the lower bounds achievable for a nonscar initial state. Note that these were computed using Eqs.~\eqref{eq: bound_k1} and ~\eqref{eq: bound_k2}, with the dimensions set to $D = 16, \; D_\alpha = 15$. This is further evidenced by the distance between $ \hat{\rho}_{T}^{(k)}$ and $ \hat{\rho}_{\text{NS-Haar}}^{(k)}$ for first and second moments ($k = 1, \;2$), shown in the blue. The line is the average over 100 iterations and the shaded areas represent the 10th-90th percentile. This distance becomes arbitrarily small in the long time limit for both $k = 1$ and $k = 2$, indicating that the models display CHSSE in the nonscar subspace.}
    \label{fig:single_qmbs}
\end{figure*}

\begin{figure}
    \centering
    \includegraphics[width=0.8\linewidth]{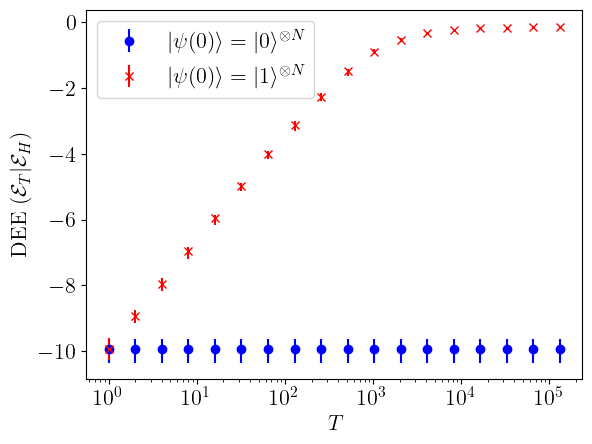}
    \caption{DEE for the dynamics generated by a brickwork circuit with a single embedded scar. Note that we have computed it for the initial scar state $\ket{0}^{\otimes N}$ (blue]) as well as a state outside the scar subspace  $\ket{1}^{\otimes N}$ (red). We use a single instance of $\{ \hat{\mathbb{U}}^{(A)}, \hat{\mathbb{U}}^{(B)}\}$. The DEE is computed using 100 different instances of sets of reference states $\{\ket{\psi_j} \}$, with $M \approx 1000$ states in each set. The error bars represent the spread between minimum and maximum values, and the crosses/dots display the mean. Since the scar state $\ket{0}^{\otimes N}$ is unaffected by the dynamics the DEE stays constant, close to the minimum value of $-\log _2 M \approx -10$ (see Appendix \ref{app: ensemble entropy}). However, DEE for the nonscar initial state $\ket{1}^{\otimes N}$ is saturating around $-0.1$, due to inability to sample any overlaps with the scar state.}
    \label{fig:ss_ensemble_entropy}
\end{figure}

To investigate the effect of ETH-violation on a model which is otherwise CHSE, we first embed QMBS in the brickwork circuit model described in the previous section. To do this we employ the projector embedding method, which was first developed for time-independent Hamiltonian models \cite{Shi-17} and later adapted for quantum circuits \cite{Log-24a}. The key step is to modify the two-qudit gates, $\hat{U}_{e/o}^{A/B}$, so that they instead are of the form: \begin{equation} \label{eq: u_scar} \hat{U}_{e/o}^{(A/B)} = \exp\{ i \hat{P} \hat{H}_{e/o}^{(A/B)}\hat{P} \}, \end{equation} where $\hat{P}$ are two-qudit projectors, i.e., $d^2 \times d^2$ matrices for which $\hat{P}^2 = \hat{P}$. The main idea of the projectors is to ensure that the unitaries act trivially on a set of chosen target states, while evolving the states outside of this set by some complex dynamics.

The projector on the full Hilbert space, corresponding to the local projector $\hat{P}$ acting on the neighboring sites $n$ and $n+1$, is denoted $\hat{\mathbb{P}}_{n, n+1} \equiv \hat{\mathbb{I}}_{0, n-1} \otimes \hat{P} \otimes \hat{\mathbb{I}}_{n+2, N-1}$, where $\hat{\mathbb{I}}_{i, j}$ is the identity acting on all qudits in the range $i, i+1, ..., j-1, j$. The set of target states then consists of all the states that are simultaneously annihilated by all projectors:
\begin{equation} \mathcal{T} = \{ \ket{\psi_{S_i}} : \hat{\mathbb{P}}_{n,n+1} \ket{\psi_{S_i}} = 0, \forall \: n \} . \end{equation}
This protocol of embedding QMBS in a circuit model effectively creates dynamically disconnected subspaces. In case the model supports $L$ scar states, the full Hilbert space, $\mathcal{H}_F$, will consist of $L+1$ disconnected subspaces, where $L$ of those will be scar subspaces containing a single state, $\mathcal{H}_{S_i} = \{\ket{\psi _{S_i}} \bra{\psi_{S_i}} \}$, such that:
\begin{equation}
    \mathcal{H}_{F} = \mathcal{H}_{N} \oplus \bigoplus_{i = 1}^{L} \mathcal{H}_{S_i} = \mathcal{H}_{N} \oplus \mathcal{H}_{S},
\end{equation}
where we have collected all the scar states into $\mathcal{H}_S$ for brevity, and the $\mathcal{H}_N$ contains all of the nonscar states.
Note that in this specific circuit construction, all the QMBS are not interacting between themselves, but one could easily construct nonproduct QMBS by using another layer of unitaries corresponding to classical cellular automata \cite{Iad-20a}.

This embedding process will not only ensure that the scar states do not thermalize (in any framework, neither CHSE nor ETH). However, it will not impact just the states in $\mathcal{H}_{S}$, but the evolution of states in $\mathcal{H}_{N}$ as well. Since the two subspaces are disconnected by construction, the $k$-th moment of the temporal ensemble, $\hat{\rho}_T^{(k)}$, generated 
by the dynamics starting from any initial state $\ket{\phi _0} \in \mathcal{H}_N$ will not be converging to $\hat{\rho}_{\text{Haar}}^{(k)}$ in the full Hilbert space. However, since the used unitaries $\mathbb{\hat{U}^{(A/B)}}$ are quite generic, we expect it to converge to the Haar averaged state in the nonscar subspace, exhibiting CHSSE within it, as described in Sec. \ref{sec:chsse}

To demonstrate the effect of embedding scars into the system, we again consider a simple setup of $N = 4$ qubits. For the construction of local unitaries, we again generate four different Hermitian operators for each of the 100 different instances, in the same manner as in Sec. \ref{sec:bw_chse}. However, the local unitaries are modified, as described in Eq.~\eqref{eq: u_scar}. The used local projector is:
\begin{equation}
    \hat{P} = \hat{\mathbb{I}} - \ket{00}\bra{00},
\end{equation}
which ensures that any of the 2-local unitaries leaves the $\ket{00}$ state invariant. The use of this projector leads to a single scar, $\ket{S_1} = \ket{0}^{\otimes N}$. The scar state will be invariant under the action of both $\hat{\mathbb{U}}^{(A)}$ and $\hat{\mathbb{U}}^{(B)}$, whereas any other, nonoverlapping, state will undergo highly nontrivial dynamics. 

To verify this, we perform the time evolution and obtain the temporal ensemble with two different initial states, the scar state $\ket{S_1} = \ket{0}^{\otimes N}$ and a state in the nonscar subspace $\ket{\psi_{\text{NS}}} = \ket{1}^{\otimes N}$. For the former we compute the distance for the $k = 1, 2$ moments with respect to the Haar moments for the full Hilbert space. For the nonscar initial state we compute the same metrics, but we also compute the distance with respect to the the first and second moments of the Haar ensemble in the nonscar subspace, denotes as $\hat{\rho}_{\text{NS-Haar}}^{(k)}$. The results are displayed in Fig. \ref{fig:single_qmbs}.

We can clearly see the difference in the dynamics for an initial scar state and nonscar state. Since the $\ket{0}^{\otimes N}$ scar is not affected by any of the unitaries in the sequence, the two moments for the temporal ensemble ($\hat{\rho}_S^{(1)} = \ket{0}^{\otimes N} \bra{0}^{\otimes N}$, $\hat{\rho}_S^{(2)} = [\ket{0}^{\otimes N} \bra{0}^{\otimes N}]^{\otimes 2}$), as well as the distances $D_{\text{HS}}(\hat{\rho}_T^{(k)}, \hat{\rho}_{\text{Haar}}^{(k)})$, with $k = 1, \; 2$, will be constant, as demonstrated. 

For the nonscar state $\ket{1}^{\otimes N}$ we see that the generated first and second moments of the temporal ensemble, $\hat{\rho}_T^{k = 1, 2}$, are clearly getting closer to the respective moments of the Haar ensemble. However, due to the separation of the scar subspace, this initial state will be constrained to exploring only the nonscar subspace. This becomes evident when comparing it to the Haar ensembles for the two moments, $\hat{\rho}_{\text{Haar}}^{(k = 1, 2)}$, as well as the to the nonscar subspace constrained Haar ensemble moments $\hat{\rho}_{\text{NS-Haar}}^{(k = 1, 2)}$. Note that the former is constructed by using Eq.~\eqref{eq: haar_moments}, and for the latter we use Eq.~\eqref{eq: subspace-haar} with $\mathcal{K}_\alpha = \mathcal{H} \; \setminus \; \mathcal{H}_S$, where $\mathcal{H}_S = \text{span}(\{ \ket{0}^{\otimes N} \})$.

When computing the distance between the generated temporal ensemble moments, $\hat{\rho}_T^{(k)}$, and the full space Haar moments, $\hat{\rho}_H^{(k)}$, we observe that these are bounded by $D_{\text{HS}}(\hat{\rho}_{\text{Haar}}^{(k)}, \hat{\rho}_{\text{NS-Haar}}^{(k)})$, with $k = 1, 2$. These bounds can be computed exactly (see Appendix \ref{app: bounds_derivation}):

\begin{align}
    D_{HS}(\hat{\rho}_{\text{Haar}}^{(1)}, \hat{\rho}_{\mathcal{K}_\alpha-\text{Haar}}^{(1)}) = \frac{D - D_\alpha}{D_\alpha D}, \label{eq: bound_k1}
    \\
    D_{HS}(\hat{\rho}_{\text{Haar}}^{(2)}, \hat{\rho}_{\mathcal{K}_\alpha-\text{Haar}}^{(2)}) = \frac{2(D^2+D - D_\alpha^2 - D_\alpha)}{D_\alpha (D_\alpha + 1) D (D + 1)},
    \label{eq: bound_k2}
\end{align}
where $D = \text{dim} (\mathcal{H}), \; D_\alpha = \text{dim}(\mathcal{K}_\alpha)$. 
For this particular model, the $k =1$ bound is $\frac{1}{D(D-1)}$, and for $k =2 $ it is $\frac{4}{(D+1)D(D-1)}$. Since $D = 2^N$, both of these bounds are exponentially small in system size $N$ (further details are in Appendix \ref{app: bounds_derivation}).
When comparing the generated temporal ensembles to the subspace Haar ensembles directly, via $D_{\text{HS}}(\hat{\rho}_{T}^{(k)}, \hat{\rho}_{\text{NS-Haar}}^{(k)})$, we see that they become arbitrarily small, and effectively reproduce the behavior observed for a CHSE model, as shown in Fig. \ref{fig:bw_k1}. The observed results for the time evolution of $\Delta^{(1)}$ and $\Delta^{(2)}$ can be directly related to the commutant algebra of the constructed model, which is described in more detail in Appendix  \ref{sec:commutant_algebras}. We also compute the DEE starting in the scar as well as a nonscar state, displayed in Fig. \ref{fig:ss_ensemble_entropy}. The DEE is independent of temporal sample size for the initial scar state. However, starting in a nonscar state the DEE is converging to $\sim -0.1$ for increasing $T$. 

The presence of CHSSE in the nonscar subspace also directly implies that the constructed states are forming $t$-designs within that subspace \cite{Nak-21, Pil-23a}. However, as demonstrated by the numerical results, the convergence rates to form $t$-designs are rather slow. 

Here we have considered the example of a single scar, but the results can be easily generalized to different numbers of scars. We provide numerical evidence for these examples in Appendix \ref{app:more_scars}.

\begin{figure*} 
    \centering
    \includegraphics[width=0.49\linewidth]{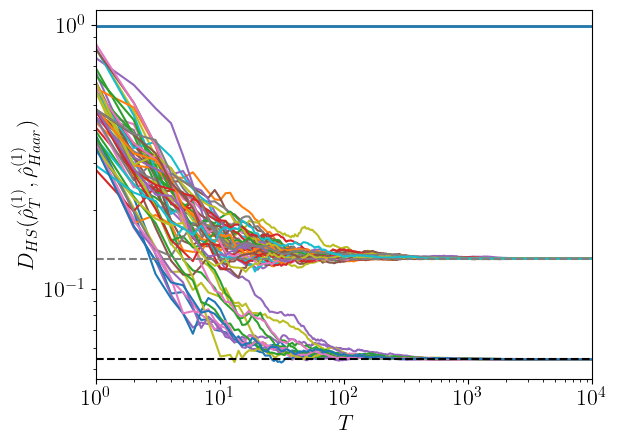}
    \includegraphics[width=0.49\linewidth]{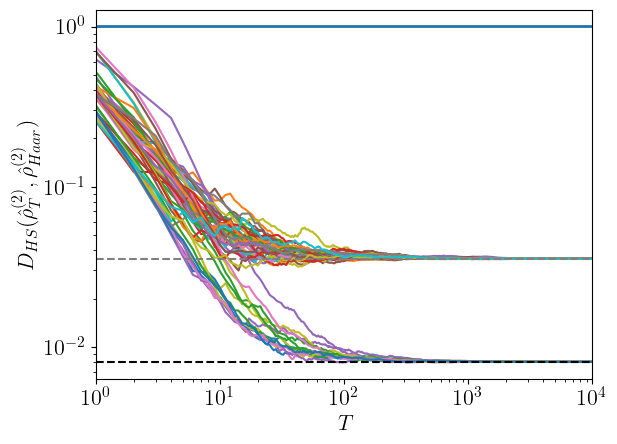}
    \includegraphics[width=0.49\linewidth]{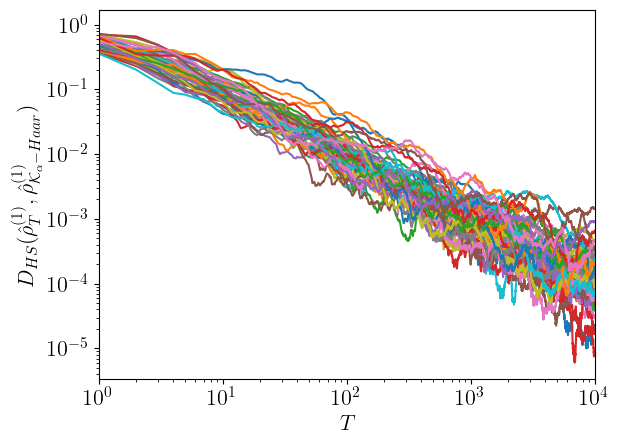}
    \includegraphics[width=0.49\linewidth]{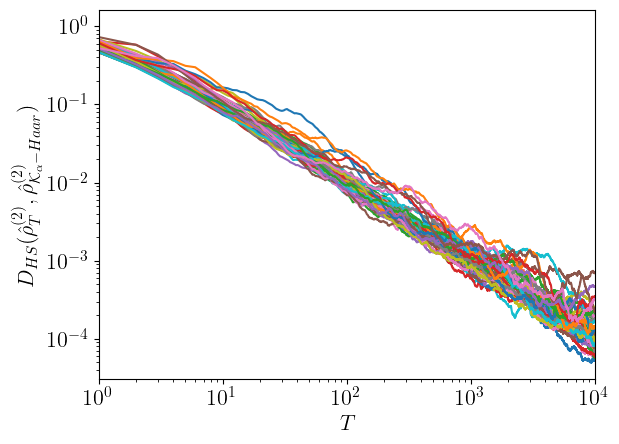}
    \caption{Top: Hilbert-Schmidt distance between the first (left) and second (right) moments of the Haar ensemble, and the temporal ensemble generated by the Fibonacci layered pair-flip model, with $N = 4$, $d = 3$. We test all possible initial computational basis states to generate the temporal ensembles with a single instance of $\{\hat{\mathbb{U}}^{(A)}, \hat{\mathbb{U}}^{(B)} \}$. For the frozen states, the distance to the first Haar moment stays constant, at around $\mathcal{O}(1)$. The $42$ states belonging to $7$-dimensional subspaces decay and eventually saturate at the distance between the full Hilbert space Haar moments, and the Haar moments in a $7$-dimensional subspace (gray dashed line). The states belonging to the largest, $15$-dimensional Krylov subspace, however, saturate at a lower bound (black dashed line), due to the larger subspace. Both of these bounds are computed using Eqs.~\eqref{eq: bound_k1} and ~\eqref{eq: bound_k2}, with $D = 81$, $D_\alpha = 15$ (black) and $D_\alpha = 7$ (gray). Bottom: Hilbert-Schmidt distance the temporal ensemble and the subspace-restricted Haar ensemble  for the first (left) and second (right) moments. We test all initial computational basis states, apart from the frozen ones. All of the initial states are converging to the $\hat{\rho}^{(k)}_{\mathcal{K}_{\alpha}}$ in their respective subspaces, thus displaying CHSSE in them.}
    \label{fig:hsf_moments_comparison}
\end{figure*}

\section{CHSSE through Hilbert Space Fragmentation}\label{sec:fragmentation}

\begin{figure}
    \centering
    \includegraphics[width=\linewidth]{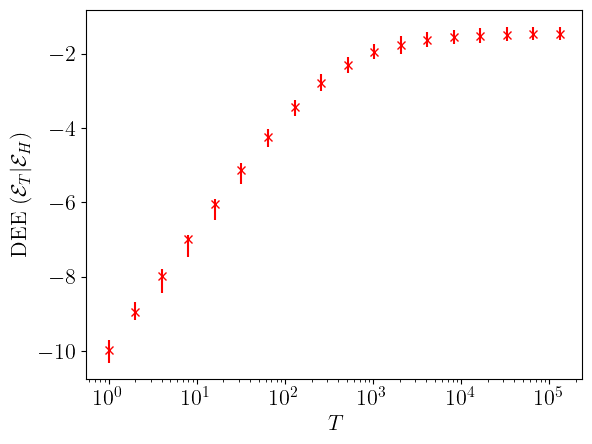}
    \caption{Discretized ensemble entropy for the dynamics generated by a brickwork circuit with the pair-flip unitary. Note that we have computed it for the initial state $\ket{0}^{\otimes N}$, which belongs to the largest, $15$-dimensional subspace. We use a single instance of $\{ \hat{\mathbb{U}}^{(A)}, \hat{\mathbb{U}}^{(B)}\}$. The DEE is computed using 100 different instances of sets of reference states $\{\ket{\psi_j} \}$, with $M \approx 1000$ states in each set. The lines represent the spread between minimum and maximum values, and the crosses display the mean. In contrast to a generic brickwork circuit, the DEE in a fragmented model saturates well below $0$, demonstrating the absence of ergodicity over the full Hilbert space.}
    \label{fig:hsf_ensemble_entropy}
\end{figure}


Another known mechanism of ETH-violation is HSF. Here we modify the model presented in Sec. \ref{sec:model} such as to observe \emph{strong} Hilbert space fragmentation. In the distinction between weak and strong violation of ETH discussed in the introduction, this mechanism corresponds to strong ETH-violation, due to exponentially many vanishingly small subspaces.

A known model that displays strong Hilbert space fragmentation for local Hilbert space dimension $d \geq 3$ is the so-called pair-flip model \cite{Cah-18a}, which was originally studied as a Hamiltonian model, but can also be implemented as a gate model. In this picture, the $2$-local unitaries have the form:

\begin{align}
    \hat{U}^{\text{PF}} = \sum_{a, b} U^{\text{PF}}_{ab} \ket{aa} \bra{bb} + \sum_{a \neq b} e^{i \phi _{ab}} \ket{ab} \bra{ab},
    \label{eq:pf_unitary}
\end{align}
where $U^{\text{PF}}_{ab}$ are elements of a Haar random unitary drawn from $U(d)$ and $\phi _{ab}$ are random phases drawn uniformly from $[0, 2\pi )$. Effectively, this model has the constraint that two neighboring sites can change their state if and only if they are in the same state, as seen in the first term. The second term implements an arbitrary phase on pairs of states where two sites differ, to break any undesirable symmetries. 

The condition that states are allowed to flip only if one of their neighbors is in the same state imposes a strong constraint on what parts of the Hilbert space are available to any initial state. Suppose we have a chain of $N$ qudits with local Hilbert space dimension $d$. If the chain is initialised in a product state where none of the neighboring sites are in the same state, it cannot transition to any other product state, but can only acquire a time dependent phase, and the dynamics becomes trivial. The number of such states is $d (d-1)^{N - 1}$, i.e., there are $d (d-1)^{N - 1}$ subspaces of dimension 1. Counting the total number of subspaces and their respective dimensions is highly nontrivial and we refer the reader to Refs. \cite{Han-24, Mou-22a, Cah-18a}. It can be shown that the number of disconnected subspaces in the pair-flip model with $d>2$ is:
\begin{align}
    \mathcal{N}_{\mathcal{K}}(N, d) = \frac{(d - 1)^{N + 1} - 1}{d - 2},
\end{align}
which is exponential in system size $N$, but still exponentially smaller than the total Hilbert space dimension $D = d^N$. Importantly, it can also be shown that the size of the largest subspace for even number of sites $N$ is:
\begin{equation}
    |\mathcal{K}_0 (N, d) | = d^N (1 + \frac{1}{2} \sum_{j = 1}^{N / 2} d^{-2j } \binom{1/2}{j} (-1)^j (2 \sqrt{d-1}) ^{2j}),
    \label{eq: hsf_max_subspace_dim}
\end{equation}
which is an exponentially vanishing fraction of the total Hilbert space, i.e. $\lim_{N \rightarrow \infty} (|\mathcal{K}_0(N, d)| / d^N) = 0$. Since the ratio of the dimension of the largest Krylov subsector versus the dimension of the full Hilbert space tends to 0, this is a form of \emph{strong} Hilbert space fragmentation. 

As seen in the discussion about a scarred system in Sec. \ref{sec:scars}, the scar subspace remained stationary, while a state in the nonscar subspace was displaying CHSSE, exploring all of the available Hilbert space. This gives us good intuition on the consequences of fragmentation. In the case of fragmented models, set up in way analogous to universal models with CHSE, we expect to see CHSSE for each of the initial states in their Krylov subspaces. To verify this we perform numerics analogous to the ones performed for models in Secs. \ref{sec:bw_chse} and \ref{sec:scars}.

We consider a system of $N = 4$ qutrits, $d = 3$. For these parameters, due to the constraints of the pair-flip model \cite{Han-24}, we expect $24$ frozen states, six distinct $7$-dimensional subspaces, and one $15$-dimensional subspace. The system undergoes dynamics defined by applying two global unitaries, $\hat{\mathbb{U}}^{(A)}$ and $\hat{\mathbb{U}}^{(B)}$. These are defined as a combination of an even and odd layers of 2-local unitaries as defined in Eq.~\eqref{eq:U_brick}. All of the individual local unitaries have the form as displayed in Eq.~\eqref{eq:pf_unitary}, but with different values for the free parameters $\{  U_{ab}^{\text{PF}}, \phi_{ab}\}$ for each of the four bricks $\{ \hat{U}_e^{(A)}, \hat{U}_o^{(A)}, \hat{U}_e^{(B)}, \hat{U}_o^{(B)} \}$ in the circuit. 

We first construct the temporal ensembles for all $81$ distinct initial computational basis states, and compare the first and second moments to the Haar ensemble, as shown in Fig. \ref{fig:hsf_moments_comparison}. Analogously to the QMBS state, the frozen states lead to a temporal ensemble for which all moments stay at a distance of $\mathcal{O}(1)$ with respect to the Haar ensemble. However, the rest of the computational basis states exhibit CHSSE in their respective subspaces, as demonstrated by their convergence to the bounds.

Note that for the largest subspace in this HSF model, the $k = 1$ bound reduces to $\frac{2}{D_\alpha}$, and the $k = 2$ bound to $\frac{2}{D_\alpha (D_\alpha + 1)}$, in the large $N$ limit (see Appendix \ref{app: bounds_derivation}). Hence, both of these are decreasing with the system size $N$, but at a slower rate than the bounds for the QMBS model.

Similarly, the fragmentation of the Hilbert space can also be observed in the DEE as a function of time, displayed in Fig. \ref{fig:hsf_ensemble_entropy}, where we only used the initial state $\ket{0}^{\otimes N}$, which belongs to the largest Krylov subspace. The DEE is converging to approximately $-2$, as expected due to the inability to sample the full Hilbert space.

\section{Role of symmetries}\label{sec:symmetries}

\begin{figure} 
    \centering
    \includegraphics[width=\linewidth]{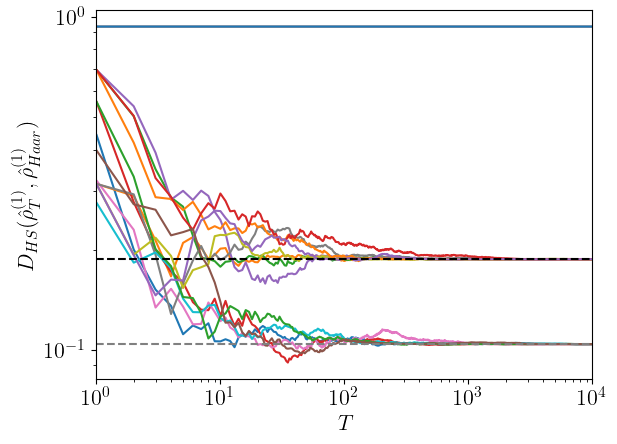}
    \includegraphics[width=\linewidth]{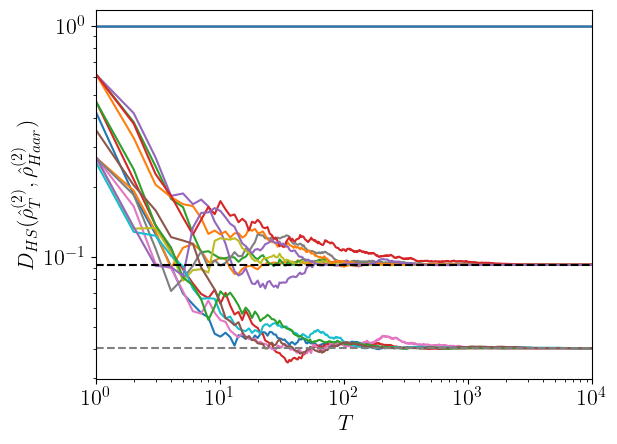}
    \caption{Hilbert-Schmidt distance between the first (top) and second (bottom) moments of the Haar ensemble, and the temporal ensemble generated by the Fibonacci layered pair-flip model, with $N = 4$, $d = 2$, for a single instance of $\{\hat{\mathbb{U}}^{(A)}, \hat{\mathbb{U}}^{(B)} \}$. We test all possible initial product states to generate the temporal ensembles. For the two invariant states, the distance to the first Haar moment stays constant, at $\mathcal{O}(1)$. The eight states belonging to four-dimensional subspaces decay and eventually saturate at the distance between the full Hilbert space Haar moments, and the Haar moments in a four-dimensional subspace (black dashed line). The states belonging to the largest, six-dimensional subspace saturate at a lower bound (grey dashed line). Both of these bounds are computed using Eq.~\eqref{eq: bound_k1} and Eq.~\eqref{eq: bound_k2}, with $D = 16$, $D_\alpha = 4$ (black) and $D_\alpha = 6$ (gray).}
    \label{fig:sym_moments_comparison}
\end{figure}

\begin{figure} 
    \centering
    \includegraphics[width=0.85\linewidth]{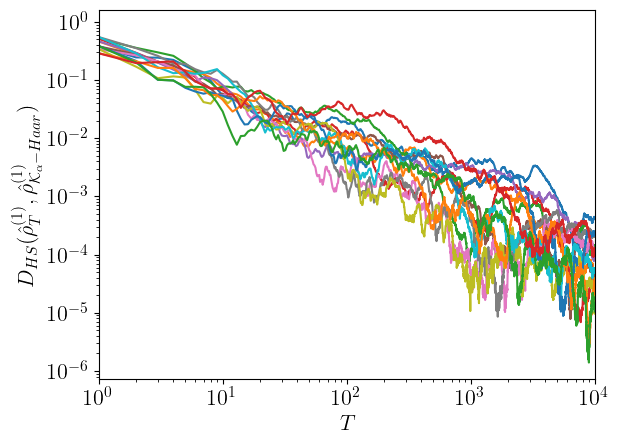}
    \includegraphics[width=0.85\linewidth]{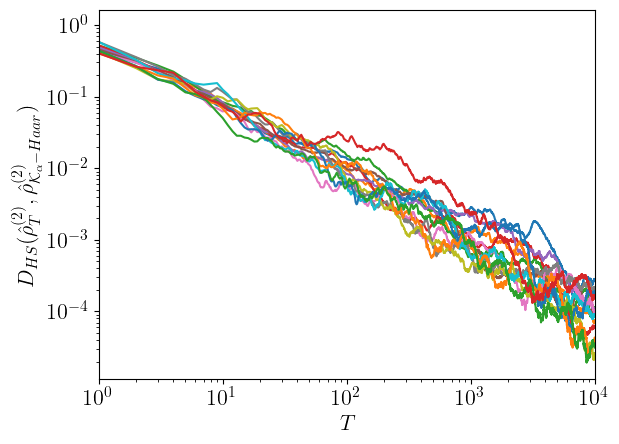}
    \caption{Hilbert-Schmidt distance the temporal ensemble and the subspace-restricted Haar ensemble for the first (top) and second (bottom) moments. We test all possible initial computational basis states, apart from the frozen ones. All of the initial states are converging to the $\hat{\rho}^{(k)}_{\mathcal{K}_{\alpha}-\text{Haar}}$ in their respective subspaces, thus displaying CHSSE in them.}
    \label{fig:sym_moments_comparison_subspace}
\end{figure}

\begin{figure}
    \centering
    \includegraphics[width=0.8\linewidth]{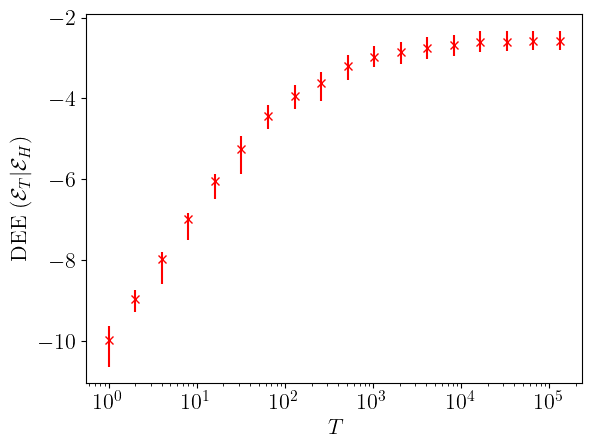}
    \caption{ Discretized ensemble entropy for the dynamics generated by a brickwork circuit with a $U(1)$ symmetry corresponding to the staggered magnetization. Note that we have computed it for the initial state $\ket{0}^{\otimes N}$, belonging to the largest symmetry subspace, and for a single instance of $\{ \hat{\mathbb{U}}^{(A)}, \hat{\mathbb{U}}^{(B)}\}$. We compute the DEE using 100 different instances of sets of reference states $\{\ket{\psi_j} \}$, with $M \approx 1000$ states in each set. The lines represent the spread between minimum and maximum values, and the crosses display the mean. We notice the DEE saturating well below the $0$ bound expected for completely Hilbert space ergodic dynamics.}
    \label{fig:sym_ensemble_entropy}
\end{figure}

In the ETH framework, the presence of a few local symmetries in a model is generally not classified as an ETH-violation mechanism, as we still observe thermal behavior within the symmetry subsectors. However, to complete the picture, we also demonstrate the effect of symmetries on an otherwise CHSE model. We again use the 2-local unitaries as defined in Eq. \eqref{eq:pf_unitary}, but in a system with $d = 2$. As any operator can only flip neighboring sites from $\ket{00}$ to $\ket{11}$ (and vice versa), one can deduce that in this case the model will have a global $U(1)$ symmetry, corresponding to the staggered magnetization, $\hat{M}_{S} = \sum_{j = 1}^{N} (-1)^j \ket{\alpha}_j\bra{\alpha}_j$. Therefore, for system size $N$, we will get a splitting into $N + 1$ subspaces, with dimensions ${N \choose j}$, where $j = 0, 1, ..., N$ labels the different subspaces. We perform similar numerical simulations as for the models in Secs. \ref{sec:scars} and \ref{sec:fragmentation}, with $N = 4$ and up to $T = 10^4$. The results are displayed in Figs. \ref{fig:sym_moments_comparison} and \ref{fig:sym_moments_comparison_subspace}. This effect can also be observed in the DEE as a function of $T$ for a state belonging to the largest subspace, as shown in Fig. \ref{fig:sym_ensemble_entropy}. The DEE saturates to a bound set by the size of the largest subspace in the long $T$ limit. Additional results are presented in Appendix  \ref{app:symmetries}. Similarly to the fragmented case, we observe almost perfect convergence to the Haar moments in symmetry subspaces. This indicates that starting in any initial state belonging to a symmetry subsector, we will form $t$ designs within that subspace for sufficiently long times.

\section{Conclusion}\label{sec:conclusion}

Complete Hilbert space ergodicity has extended the notion of ergodicity from classical to certain quantum systems undergoing aperiodic dynamics. It can thus be seen as complementary to the ETH, which has been conceptually crucial for our understanding of thermalization in Hamiltonian and Floquet systems. Through many experiments and numerical demonstrations, different ETH violating mechanisms have been identified, including QMBS and HSF.

In this work, we presented the effects of these ergodicity breaking mechanisms on systems which otherwise display CHSE. In both scarred and fragmented models, the presence of nonlocal conserved quantities leads to a decoupling between subspaces of the full Hilbert space. Thus, different initial states are constrained to their respective subspaces, and we cannot observe CHSE in the full Hilbert space, but we do observe CHSE in the subspaces. The scarring and fragmentation mechanisms in the considered aperiodic models may also be detected via the same measures that have been used to verify their presence in Hamiltonian and Floquet models, such as the growth of bipartite entanglement entropy and the decay of the autocorrelator.

The presented work opens up further avenues and questions for future research. We observe that one can construct Haar ensembles in some designated subspaces, This protocol could be adapted to construct them in decoherence free subspaces, akin to the prescription described in Ref. \cite{Wan-24a}. The constructed $t$-designs could have practical applications in fault tolerant quantum computers. Unfortunately, we observe that the temporal ensembles approximate Haar ensembles only at late times. Hence, it remains an important open problem to investigate the convergence rates for the different $k$-moments.

\begin{acknowledgments}
We acknowledge helpful conversations with Alessandro Summer, Alex Nico-Katz, Federica Maria Surace, and Nathan Keenan. J.G. and L.L. acknowledge financial support by Microsoft Ireland. J.G. is supported by a SFI-Royal Society University Research Fellowship. S.D. acknowledges the financial support of Taighde Eireann - Research Ireland under grant number 22/PATH-S/10812.
\end{acknowledgments}


\bibliography{refs}

\newpage
\clearpage

\appendix

\section{Universality of Local Hamiltonians and Brickwork Circuits} \label{app: universality}

\subsection{Universality of $k$-local Hamiltonians}

Suppose we have a quantum system with $L$ sites, with a $d$-dimensional qudit at each site, resulting in a $D$-dimensional Hilbert space $\mathcal{H}$, with $D = d^L$. Now we can apply one of two $k$-local Hamiltonians at a time, either $\hat{H}_A$ or $\hat{H}_B$, which we can write as:

\begin{equation}
    \hat{H}_{X} = \sum_{j = 0}^L \hat{h}_{X, j},
\end{equation}
where $X = A, B$, and $\hat{h}_{X, j}$ are local operators centered at site $j$. We also assume that we can apply them for some predetermined, arbitrary time $\tau _{A/B}$, which will result in the unitary $\hat{U}_{X} = e^{-i \tau _X \hat{H}_X}$. The set $\{ \hat{U}_A, \hat{U}_B \}$ will be universal (i.e. we can construct any unitary $\hat{U}_C \in \text{U}(D)$ up to arbitrary precision $\epsilon$ by applying the two universal gates in some sequence, $||\prod_{i = \{A, B \} }\hat{U}_i - \hat{U}_C|| \leq \epsilon$, where $|| \hdots ||$ denotes the trace norm), provided that \cite{Llo-95a}:
\begin{enumerate}
    \item The Hamiltonians $\hat{H}_A$ and $\hat{H}_B$ do not share any eigenstates and/or symmetries.
    \item Eigenphases of $\hat{U}_X$ are irrational multiples of $\pi$.
\end{enumerate}
We assume that the first condition is satisfied, as generic Hamiltonians (with arbitrary terms $\hat{h}_{X, j}$) will have different eigenstates and will not contain any symmetries. The second condition can be guaranteed, by using a suitable choice of the evolution times $\tau _X$. 

For each of the two Hamiltonians, we can label the eigenenergies as $E_{j}, j = 1, 2, ..., D$. Then, the eigenphases of the corresponding unitary will be $\phi _j = \tau E_j$. Now, the eigenenergies could be either rational or irrational numbers, but in either case there will exist a time $\tau$ which will make all of the $\phi _j$ irrational multiples of $\pi$. Hence, we can make the unitaries $\hat{U}_{A}, \hat{U}_B$ universal by a suitably choosing the evolution times $\tau_A$ and $\tau_B$.


\subsection{Universality of brickwork circuits}
Now, building upon the previous argument, we can construct a similar argument for the universality of unitaries corresponding to one odd and one even layer in a brickwork layer, as described in Eq.~\eqref{eq:U_brick}. Suppose we have a $2-$local Hamiltonian, $\hat{H}$, acting on an open chain of $L$ qudits. All of the terms in the Hamiltonian act only on nearest neighbors, so we can decompose it as $\hat{H} = \sum_{i = 0, 2, 4, ...}^{L-2} \hat{g}_{i, i+1}^e + \sum_{i = 1, 3, ...}^{L-3} \hat{g}_{i, i+1}^o$. Now, we let the system evolve for some time $\tau = \eta \delta$, where $\eta$ is an arbitrary number of order $\mathcal{O}(1)$ and $\delta$ is a parameter which can be arbitrarily small, and we will set it to be of the form $10^{-a}, a \in \mathbb{Z}^+$. This will result in the unitary $\hat{U} = e^{-i \tau  \hat{H}} = e^{-i \eta \delta \hat{H}}$. Now, by the argument in section above we can always set $\eta$ such that the eigenphases of $\hat{U}$ will be irrational multiples of $\pi$. 

Writing the terms in the exponent as: $\eta \delta \hat{g}_{i, i+1}^{o/e} = \tau \hat{h}_{i, i+1}^{o/e}$, we can identify the $\hat{\mathbb{U}}_t$ in Eq.~\eqref{eq:U_brick} as a trotterization of some $2$-local Hamiltonian, with a single timestep. Hence, the two unitaries are identical up to Trotter error: $\hat{U} = \hat{\mathbb{U}} + \mathcal{O}(\eta^2 \delta^2)$. Therefore we can make the error arbitrarily small, by tuning the $\delta$ parameter.

\section{Commutant algebras} \label{sec:commutant_algebras}



Given some Hamiltonian $\hat{H} = \sum_j \alpha _j \hat{h}_j$ consisting of local terms $\hat{h}_j$ (or similarly, a unitary composed of local gates $\hat{U} = \prod_j e^{i \alpha_j \hat{h}_j}$), the commutant algebra ($\mathcal{C}$) is the set of Hermitian operators $\hat{c}_k$, which commute with each individual term in the Hamiltonian $\hat{H}$:
\begin{equation}
    [\hat{h}_j, \hat{c}_k] = 0 \: \forall \: j.
\end{equation}
Together with the commutant algebra, we also define the so-called bond algebra ($\mathcal{A} \equiv \langle \langle \hat{H} \rangle \rangle$), which is the associative algebra obtained by taking linear combinations with complex coefficients of arbitrary products of all the terms in $\hat{H}$ and the identity operator $\hat{\mathbb{I}}$ (note that we can always include the identity in a given Hamiltonian by using an energy off-set, which does not change the dynamics). It is called the ``bond algebra'' since taking sums and products of the local $\hat{h}_j$ will result in longer and longer strings of operators, thus creating bonds. 

We use $\mathcal{L}(\mathcal{H})$ to denote the set of all linear operators on $\mathcal{H}$. Then $\mathcal{A}$, $\mathcal{C}$ will be subsets of $\mathcal{L}(\mathcal{H})$ and most importantly centralizers of each other \cite{Mou-22a}. Due to this fact, the irreducible representations of $\mathcal{A}, \mathcal{C}$ can be used to construct a virtual bipartition of the Hilbert space: $\mathcal{H} = \bigoplus_{\lambda} \mathcal{H}_{\lambda}^{(\mathcal{A})} \otimes \mathcal{H}_{\lambda}^{(\mathcal{C})}$. There exists a basis, in which the operators $\hat{h}_{\mathcal{A}} \in \mathcal{A}$, $\hat{h}_{\mathcal{C}} \in \mathcal{C}$ can be written as:

\begin{align} \label{eq: subspace_matrices}
    \hat{h}_{\mathcal{A}} = \bigoplus_{\lambda} (\hat{M}_{D_{\lambda}}^{\lambda} (\hat{h}_{\mathcal{A}}) \otimes \hat{\mathbb{I}}_{d_{\lambda}}),\notag \\ \hat{h}_{\mathcal{C}} = \bigoplus_{\lambda}  \hat{\mathbb{I}}_{D_{\lambda}} \otimes (\hat{M}_{d_{\lambda}}^{\lambda} (\hat{h}_{\mathcal{C}})),
\end{align}
where $D_{\lambda}$ and $d_{\lambda}$ are the representations of irreps of $\mathcal{A}$ and $\mathcal{C}$, $\hat{M}_{D_{\lambda}}^{\lambda} (\hat{h}_{\mathcal{A}}$ and $\hat{M}_{d_{\lambda}}^{\lambda} (\hat{h}_{\mathcal{C}})$ are the $D_{\lambda}$ and $d_{\lambda}$ dimensional matrix representations of the operators in the arguments. As such, this is just a matrix representation of the operators in the basis in which all the operators in $\mathcal{A}$ and $\mathcal{C}$ are block diagonal. Note that the dimensions of $\mathcal{A}, \mathcal{C}$ are the dimensions of the subspaces of matrices used in Eq. \eqref{eq: subspace_matrices}:
\begin{equation}
    \text{dim}(\mathcal{A}) = \sum_{\lambda} D_{\lambda}^2, \; \text{dim}(\mathcal{C}) = \sum_{\lambda} d_{\lambda}^2, \; \text{dim}(\mathcal{H}) = \sum_{\lambda} D_{\lambda} d_{\lambda}
\end{equation}

\subsection{Commutant algebra of the QMBS model}

In Sec. \ref{sec:scars}, all of the $2$-local unitaries have the form as written in Eq.~\eqref{eq: u_scar}. By construction, all of the target states, $\{\ket{\psi _{S_i}} \}$ commute with the $2$-local unitaries used to construct the circuit model, i.e.:

\begin{equation}
    [\hat{U}_{t, e/o}^{n, n+1}, \ket{\psi _{S_i}}  \bra{\psi _{S_i}}] = 0,
\end{equation}
where the gates have the form: $\hat{U}_{t, e/o}^{n,n+1} = \exp\{ i \hat{P}^{n,n+1} \hat{H}_{t, e/o}^{n,n+1}\hat{P}^{n,n+1} \}$, and $\ket{\psi _{S_i}}$ are the states that simultaneously annihilated by all the projectors. This immediately implies that the commutant algebra of this model consists of the projectors corresponding to each of the scar states and the identity:
\begin{equation}
    \mathcal{C}_{\text{QMBS}} = \{ \ket{\psi _{S_i}} \bra{\psi _{S_i}}, \hat{\mathbb{I}} \}.
\end{equation}
More concretely, in the case of a single scar, we will have two operators in the commutant algebra, dim$(\mathcal{C}) = 2$, implying that we have two distinct $\lambda$ values with $d_{\lambda} = 1$ for both. Note that due to the fact that the scar subspace is one-dimensional, we can immediately read of that $D_{\lambda_1} = 1$, $D_{\lambda_2} = \text{dim}(\mathcal{H}) - 1$. Hence, we will have one one-dimensional subspace corresponding to the scar state, and one $\text{dim}(\mathcal{H}) - 1$ dimensional, thermal subspace.

\subsection{Commutant algebra of the pair-flip model}

For a more in depth discussion of the pair-flip model commutant algebra see Ref. \cite{Mou-22b}, where they provide an in-depth diagrammatic model and explanation. To construct the commutant algebras, we define the following operators:
\begin{equation}
    \hat{N}_{i}^{\alpha} = \ket{\alpha}\bra{\alpha}_i, \hat{N}^{\alpha} = \sum_j (-1)^j \hat{N}_{j}^{\alpha},
\end{equation}
where we can think of $\hat{N}_i^{\alpha}$ as the ``counting operator'' determining whether the state at site $i$ is $\ket{\alpha}$ or not. It is easy to verify that $\hat{N}^{\alpha}$ commutes with the pair-flip unitaries of Eq.~\eqref{eq:pf_unitary}, thus implying the existence of a $U(1)$ symmetry due to (slightly modified) particle number conservation.

However, due to the observation that we get exponentially many subspaces, there should also exist other conserved quantities which are linearly independent of $\hat{N}^{\alpha}$. First, we will denote the ``flip'' part of the unitary as:
\begin{equation}
    \hat{F}_{ij}^{\alpha \beta} = U^{\text{PF}}_{ab} \ket{aa}_{ij} \bra{bb}_{ij}.
\end{equation}
Now we can extend to previously found commuting operators by observing that:
\begin{equation}
    [\hat{N}_i^{\sigma} \hat{N}_j^{\tau}, \hat{F}_{ij}^{\alpha \beta}] = 0, \text{for} \sigma \neq \tau, 1 \leq \alpha, \beta, \sigma, \tau \leq d,
\end{equation}
where the last two constraints are trivial, just making sure we do not overcount the operators. Now, using this expression we can construct higher order operators that commute with the pair-flip unitaries:
\begin{equation} \label{eq: pf_commuting_observables}
    \hat{N}^{\alpha_1, ..., \alpha_k} \equiv \sum_{j_1 < j_2 < ... < j_k} (-1)^{\sum_{l = 1}^k j_l} \hat{N}_{j_1}^{\alpha_1} ... \hat{N}_{j_k}^{\alpha_k}.
\end{equation}
However, note that not all of the operators encapsulated by Eq.~\eqref{eq: pf_commuting_observables} are linearly independent. For even $N$, the operators with odd $k$ can be expressed as a linear combination of the operators with even $k$. The total number of linearly independent operators in $\mathcal{C}$ is then:

\begin{equation}
    \text{dim}(\mathcal{C}) = 
    \begin{cases}
        \frac{(d-1)^{N+1} - 1}{d - 2} & \text{if } d \geq 3 \\
        N+1 & \text{if } d = 2,
    \end{cases}
\end{equation}
showing that the pair-flip model contains a regular symmetry for $d = 2$ and that it should exhibit Hilbert space fragmentation for $d > 2$. This is clearly seen by the numerical results in Secs. \ref{sec:fragmentation} and \ref{sec:symmetries}, where we observe the Hilbert subspace ergodicity in each of the disconnected subspaces.

\section{Numerical procedure for computing Discretized Ensemble Entropy} \label{app: ensemble entropy}

In practice, we will end up with a finite set of $T$ states in the temporal ensemble: $\ket{\Psi ^T _0 }, \ket{\Psi ^T _1 }, \ket{\Psi ^T _2 }, ..., \ket{\Psi ^T _{T-1} }$., from which we want to compute an estimator for the discretized ensemble entropy as written in Eq.~\eqref{eq: ens_ent}. One potential approach would be to exactly separate the Hilbert space into different sections, which are occupying the same ``volume,'' but this would be quite difficult for arbitrarily large Hilbert space dimension. Therefore, we perform
a randomized sampling procedure to estimate the probabilities, $\{p_j^T, p_j^H \}$ corresponding to the temporal and Haar ensembles: \begin{enumerate}
    \item Pick $T$, $M$ and $\epsilon$, with $\epsilon \in [0, 1)$. $T$ and $M$ should be chosen such that $T \gg M$, to minimize the fluctuations in probabilities.
    \item Sample $T$ states generated by the time evolution of chosen aperiodic drive, $\{ \ket{\Psi _i^T} \}$, as well as $T$ states from the Haar ensemble, $\{ \ket{\Psi _j^H} \}$.
    \item Sample $M$ states from Haar ensemble, $\{ \ket{\Phi _j^R} \}$, which will form a set of reference states used to estimate the probabilities.
    \item Remove states from the second set if their overlap is large with any other state, i.e. if $|\langle \Phi _j | \Phi _k \rangle| > 1 - \epsilon$, resulting in $M'$ states. This ``filtering'' step is performed to ensure numerical stability.
    \item Now, the filtered second set will form a reference set for the two sets of states $\{ \ket{\Psi _i^T} \}$ and $\{ \ket{\Psi _i^H} \}$. We will ``bin'' the states from these two sets into $M'$ bins, depending on which state they have the largest overlap with. That is, we will assign the $i$-th state state from the temporal (Haar) ensemble to the bin $j$ which maximises the overlap $|\langle \Psi _i^{T (H)} | \Phi _j \rangle|$, giving two sets of counts, $n_j^T$ and $n_j^H$.
    \item This will give us two sets of probabilities for the temporal and discretized haar ensembles as: $p_j^T = n_j^T / \sum_{j = 1}^{M'} n_j^T$ and $p_j^H = n_j^H / \sum_{j = 1}^{M'} n_j^H$. Using these in Eq.~\eqref{eq: DEE} we obtain the discretized ensemble entropy estimate.
 \end{enumerate}


Note that the discretized ensemble entropy in Eq.~\eqref{eq: DEE} is a good proxy for the ensemble entropy in Eq.~\eqref{eq: ens_ent}, but there are some minor distinctions. Both quantities are upper bounded by $0$, but the lower bounds differ. The latter is not bounded from below and can in principle tend to $-\infty$, whereas the former has a finite lower bound. The lower bound can be reached in the case of a scar, which remains unchanged throughout the time evolution and thus always has maximum overlap with the same reference state, $\ket{\Phi _{k'}}$, for some $k'$. This gives the minimum attainable value for DEE: 
\begin{align}
    \sum_{j = 1}^{M'} \delta_{j, k'} \log_2 (\frac{\delta_{j, k'}}{p_j^H}) \notag \\
    \approx \sum_{j = 1}^{M'} \delta_{j, k'} \log_2 (\frac{\delta_{j, k'}}{1/M'}) = -\log_2(M'),
\end{align}
where we used the approximation that all the probabilities associated with the reference ensemble are equally likely. Note that the same argument can be extended to the case where we are sampling Haar states within a larger subspace, as for example in the case of symmetries or fragmentation. In case the initial state is constrained to a $B$-dimensional subspace, the dynamics will generate states in the same $~\frac{B}{D}M = K$ bins. Hence, in the large $T$ limit the DEE will converge to approximately:
\begin{equation}
    \sum_{j = 1}^{K} \frac{1}{K} \log_2(\frac{M}{K}),
\end{equation}
which agrees with the large $T$ limit in Figs. \ref{fig:hsf_ensemble_entropy} and \ref{fig:sym_ensemble_entropy}.

Note that, for a finite set of bins $M'$, one can construct dynamics which are not really Hilbert space ergodic, but are just cyclically orbiting between some states belonging to each of the $M'$ regions of Hilbert space. Therefore, we also perform averaging over different choices of the reference states $\{ \ket{\Psi ^R _j}$. In case the estimates of DEE for all different reference sets agree, one can infer that the dynamics indeed exhibits Hilbert (sub-)space ergodicity.

\section{Derivation of distances between full Hilbert space and subspace Haar moments} \label{app: bounds_derivation}

First we will compute the result in Eq.~\eqref{eq: bound_k1}, the Hilbert-Schmidt distance between $\hat{\rho}^{(k = 1)}_{\text{Haar}}$ and $\hat{\rho}^{(k = 1)}_{\mathcal{K}_\alpha \textendash \;\text{Haar}}$, using Eqs.~\eqref{eq: haar_moments} and \eqref{eq: subspace-haar}. First we note that for $k = 1$ the moments have a particularly simple form. They are both proportional to the identity, with the subspace Haar moment constrained to its subspace:
\begin{eqnarray}
    \hat{\rho}_{\text{Haar}}^{(k = 1)} = \frac{1}{D} \hat{\mathbb{I}}_{D}, \; \hat{\rho}_{\mathcal{K}_\alpha -\text{Haar}}^{(k = 1)} = \frac{1}{D_\alpha} (\hat{\mathbb{I}}_{D_\alpha} + \hat{\mathbb{O}}_{D-D_\alpha}),
\end{eqnarray}
where $D = \text{dim}(\mathcal{H}), D_{\alpha} = \text{dim}(\mathcal{K}_{\alpha})$, $\hat{\mathbb{I}}_M$ is the $M$ by $M$ identity matrix, and $\hat{\mathbb{O}}_M$ is the $M$ by $M$ zero matrix. The difference between the two is then:
\begin{eqnarray}
    \hat{\delta}^{(1)} = \hat{\rho}_{\text{Haar}}^{(k = 1)} - \hat{\rho}_{\mathcal{K}_\alpha-\text{Haar}}^{(k = 1)} \notag \\ =  \frac{1}{D}\hat{\mathbb{I}}_{D - D_\alpha} \oplus (\frac{1}{D} - \frac{1}{D_\alpha}) \hat{\mathbb{I}}_{D_\alpha}, \notag
\end{eqnarray}
which is diagonal and Hermitian. Therefore:
\begin{eqnarray}
    (\hat{\delta}^{(1)})^{\dagger} \hat{\delta}^{(1)} = \hat{\delta}^{(1)}\hat{\delta}^{(1)} \notag \\ = \frac{1}{D^2}\hat{\mathbb{I}}_{D - D_\alpha} \oplus (\frac{1}{D} - \frac{1}{D_\alpha})^2 \hat{\mathbb{I}}_{D_\alpha}. \notag
\end{eqnarray}
Finally, using the obtained expression we get that:
\begin{eqnarray}
    D_{HS}(\hat{\rho}_{\text{Haar}}^{(k = 1)}, \hat{\rho}_{\mathcal{K}_\alpha \textendash\;\text{Haar}}^{(k = 1)}) = \text{Tr}[(\hat{\delta}^{(1)})^{\dagger} \hat{\delta}^{(1)}] \notag \\ = \text{Tr}[\frac{1}{D^2}\hat{\mathbb{I}}_{D - D_\alpha} \oplus (\frac{1}{D} - \frac{1}{D_\alpha})^2 \hat{\mathbb{I}}_{D_\alpha}] \notag \\ = \frac{D - D_\alpha}{D^2} + D_\alpha (\frac{1}{D} - \frac{1}{D_\alpha})^2 \notag \\ = \frac{1}{D_\alpha} - \frac{1}{D} = \frac{D - D_\alpha}{D_\alpha D}.
\end{eqnarray}
Note that for a QMBS model with a single scar we have that $D_\alpha = D - 1$ for the nonscar subspace. Therefore, in this case the bound reduces to $\frac{1}{D(D+1)}$. Since the total Hilbert space dimension is $d^N$, this bound is exponentially vanishing with system size. However, it is important to note that the notions of CHSE and CHSSE is not well-defined in the thermodynamic limit. This is due to the fact that the Haar measure, and consequently its moments, are not defined for an infinitely dimensional Hilbert space.

For our HSF model, the bound is also decreasing with system size, but at a much lower rate. First we note that the full Hilbert space dimension is growing more quickly with system size than the dimension of the largest subspace [due to Eq.~\eqref{eq: hsf_max_subspace_dim}]. Hence, for large $N$, the bound reduces to $1/D_\alpha = 1/|\mathcal{K}_0|$ for the largest subspace. 

Performing the same calculation for $k = 2$ is slightly more involved. In this case the moments are proportional to the sum of the identity and a generalized swap operator acting on two $D_{(\alpha)}$-dimensional Hilbert spaces for the full space (subspace) Haar measures:

\begin{eqnarray}
    \hat{\rho}^{(k = 2)}_{\text{Haar}} = \frac{\hat{\mathbb{I}}_{D^2} + \sum_{\beta_1, \beta_2 = 1}^{D} \ket{\beta_1, \beta_2} \bra{\beta_2, \beta _1}}{D(D + 1)} \\
,     \hat{\rho}^{(k = 2)}_{\mathcal{K}_\alpha - \text{Haar}} = \frac{\hat{\mathbb{I}}_{D_\alpha^2} + \sum_{\beta_{1,2}= 1}^{D_\alpha} \ket{\beta_1, \beta_2} \bra{\beta_2, \beta _1}}{D_\alpha(D_\alpha + 1)},
\end{eqnarray}
where we have omitted the $\hat{\mathbb{O}}$ part in $\hat{\rho}^{(k = 2)}_{\mathcal{K}_\alpha \textendash \; \text{Haar}}$ for brevity, and without loss of generality ordered the $\beta_{1, 2}$ indices such that $\ket{\beta_i} \in \mathcal{K}_\alpha,\; \text{if} \; \beta_i \leq D_{\alpha} $ and $\ket{\beta_i} \notin \mathcal{K}_\alpha,\; \text{if} \; \beta_i > D_{\alpha} $. To simplify the calculation, define: $x = \frac{1}{D(D+1)}$ and $y = \frac{1}{D(D+1)} - \frac{1}{D_{\alpha}(D_{\alpha}+1)}$. First we note that the second contribution to the moments is the generalised swap operator acting on two $D_{(\alpha)}$ dimensional qudits and as such has a diagonal part where $\beta_1 = \beta _2$ and an off-diagonal part for $\beta_1 \neq \beta_2$. Second, when taking the difference between the two moments, the subspace Haar moment will only affect the elements where both $\beta_1$ and $\beta_2$ belong to $\mathcal{K}_{\alpha}$. Using these, we obtain:

\begin{eqnarray}
    \hat{\delta}^{(2)} = \hat{\rho}^{(k = 2)}_{\text{Haar}} - \hat{\rho}^{(k = 2)}_{\mathcal{K}_\alpha - \text{Haar}} \notag \\
    = x\sum_{\substack{\beta_1 \notin \mathcal{K}_{\alpha} \\ \text{or} \; \beta_2 \notin \mathcal{K}_{\alpha}}} \ket{\beta_1, \beta_2} \bra{\beta_1, \beta _2} + x \sum_{\beta_1 \notin \mathcal{K}_{\alpha}} \ket{\beta_1, \beta_1} \bra{\beta_1, \beta _1} \notag \\
    + x\sum_{\substack{\beta_1 \notin \mathcal{K}_{\alpha} \\ \text{or} \; \beta_2 \notin \mathcal{K}_{\alpha}, \\ \beta_1 \neq \beta_2 }} \ket{\beta_1, \beta_2} \bra{\beta_2, \beta _1} + y \sum_{\beta_{1, 2} \in \mathcal{K}_{\alpha}} \ket{\beta_1, \beta_2} \bra{\beta_1, \beta _2} \notag \\ + y \sum_{\beta_1 \in \mathcal{K}_{\alpha}} \ket{\beta_1, \beta_1} \bra{\beta_1, \beta _1} + y \sum_{\substack{\beta_{1, 2} \in \mathcal{K}_{\alpha}  \\ \beta_1 \neq \beta_2 }} \ket{\beta_1, \beta_2} \bra{\beta_2, \beta _1}, \notag
\end{eqnarray}
which is Hermitian. Now:
\begin{eqnarray}
    (\hat{\delta}^{(2)})^{\dagger} \hat{\delta}^{(2)} = \hat{\delta}^{(2)} \hat{\delta}^{(2)} \notag \\
    = x^2 \sum_{\substack{\beta_1 \notin \mathcal{K}_{\alpha} \\ \text{or} \; \beta_2 \notin \mathcal{K}_{\alpha}}} \ket{\beta_1, \beta_2} \bra{\beta_1, \beta _2} + 3x^2 \sum_{\beta_1 \notin \mathcal{K}_{\alpha}} \ket{\beta_1, \beta_1} \bra{\beta_1, \beta _1} \notag \\
    + x^2 \sum_{\substack{\beta_1 \notin \mathcal{K}_{\alpha} \\ \text{or} \; \beta_2 \notin \mathcal{K}_{\alpha}, \\ \beta_1 \neq \beta_2 }} \ket{\beta_1, \beta_2} \bra{\beta_2, \beta _1} + y^2 \sum_{\beta_{1, 2} \in \mathcal{K}_{\alpha}} \ket{\beta_1, \beta_2} \bra{\beta_1, \beta _2} \notag \\ + 3y^2 \sum_{\beta_1 \in \mathcal{K}_{\alpha}} \ket{\beta_1, \beta_1} \bra{\beta_1, \beta _1} + y^2 \sum_{\substack{\beta_{1, 2} \in \mathcal{K}_{\alpha}  \\ \beta_1 \neq \beta_2 }} \ket{\beta_1, \beta_2} \bra{\beta_2, \beta _1}, \notag \\ + \text{traceless terms}. \notag 
\end{eqnarray}
Finally, we can compute the distance:
\begin{eqnarray}
    D_{HS}(\hat{\rho}^{(k = 2)}_{\text{Haar}}, \; \hat{\rho}^{(k = 2)}_{\mathcal{K}_\alpha - \text{Haar}}) = \text{Tr}[(\hat{\delta}^{(2)})^{\dagger} \hat{\delta}^{(2)}]\notag \\
    = x^2(D^2 - D_\alpha^2) + 3x^2(D-D_\alpha)\notag \\ + x^2(D^2 - D_\alpha^2 - D + D_\alpha) + y^2D_\alpha^2 \notag \\ + 3y^2 D_\alpha + y^2D_\alpha(D_\alpha - 1) \notag \\ = \frac{2}{D_\alpha (D_\alpha  + 1)} - \frac{2}{D (D  + 1)}  \\ = \frac{2D_\alpha (D_\alpha  + 1) - 2 D (D  + 1)}{D_\alpha (D_\alpha  + 1)D (D  + 1)}.
\end{eqnarray}
As displayed in Eq.~\eqref{eq: bound_k2}. Similarly to the $k = 1$ bound, we can see that this bound will decrease with system size for both the QMBS and HSF models. For the former, $D_\alpha = D-1$ in the case of a single scar, and therefore the $k = 2$ bound reduces to $\frac{4}{(D+1)D(D-1)}$. Since $D = d^N$, this bound is exponentially vanishing in system size. For the HSF model, however, this bound reduces to $\approx \frac{2}{D_\alpha (D_\alpha + 1)}$ in the large $N$ limit where $D \gg D_\alpha$. Hence, this bound is also decreasing with system size $N$, but at a much slower rate than the bound for the QMBS model.

\section{Growth of bipartite entanglement entropy for the presented models} \label{app:bipartite_entanglement_entropy}

Bipartite entanglement entropy is frequently used quantity to detect the presence of ETH-violating eigenstates. For a chain of $N$ qudits, the bipartition entanglement entropy is defined as:
\begin{equation}
    S^{vN}(\ket{\psi(t)}) = -\text{Tr}[\hat{\rho}_A(t) \text{log}(\hat{\rho}_A(t))],
\end{equation}
where $\hat{\rho}_A(t) = \Tr_{0,(N/2)-1} \ket{\psi(t)}\bra{\psi(t)}$ is the reduced density matrix of the time evolved state obtained by tracing out the first $N / 2$ qudits of the system. In the case of an ergodic system with scars, and starting in a nonscar initial state $\ket{\psi _0}$, we expect the bipartition entanglement entropy to approach the Page value \cite{Pag-93a}. 

\begin{figure}
    \centering
    \includegraphics[width=1\linewidth]{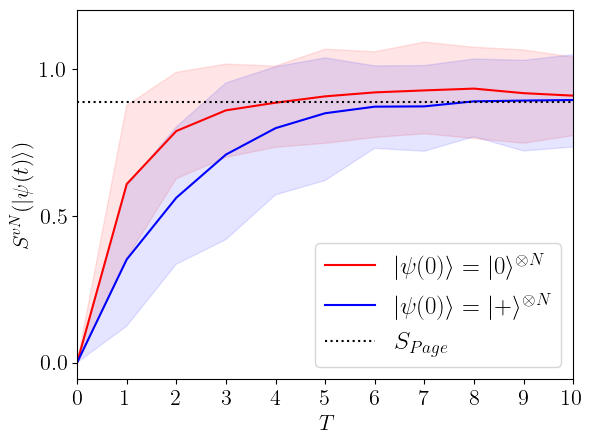}
    \caption{Growth of bipartition entanglement entropy for the generic brickwork model, starting from two  initial states: $\ket{\psi (0)} = \ket{0}^{\otimes N}$ and $\ket{\psi (0)} = \ket{+}^{\otimes N}$, for $N = 6, \; d= 2$. The results are averaged over 100 different instances, with the lines corresponding to the average distances, and the shaded areas represent the 10th-90th percentile range. Both initial states eventually saturate around $S_{\text{Page}} = \frac{1}{2}N \text{ln}(d) - \frac{1}{2}$.}
    \label{fig:bw_entanglement_entropy}
\end{figure}

We first compute the growth of bipartition entanglement entropy for two the generic brickwork model, discussed in Sec. \ref{sec:bw_chse}, with the results displayed in Fig. \ref{fig:bw_entanglement_entropy}. For both chosen initial states, $\ket{0}^{\otimes N}$ and $\ket{+}^{\otimes N}$, the bipartite entanglement initially grows quickly and reaches the Page entropy, $S_{\text{Page}} = \frac{1}{2}N \text{ln}(d) - \frac{1}{2}$.

\begin{figure}
    \centering
    \includegraphics[width=0.98\linewidth]{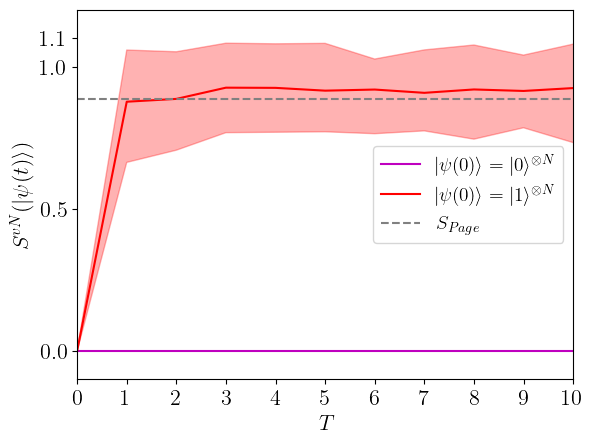}
    \caption{Time evolution of bipartite entanglement entropy for the initial scar state $\ket{S_1}$ and nonscar state $\ket{1}^{\otimes N}$, for QMBS model with $N = 6, \; d =2$. For the scar state (violet) the entanglement entropy stays at 0, which is expected for a product state. But for the nonscar state (red), the entanglement entropy instantaneously grows to the Page entropy, $S_{\text{Page}} = \frac{1}{2}N \text{ln}(d) - \frac{1}{2}$, which is the expected entanglement entropy for a Haar random state. Note that the line represents the average over the 100 iterations and the shaded region represents the 10th-90th percentiles.}
    \label{fig:single_qmbs_entropy}
\end{figure}

The results corresponding to the scarred model, presented in Sec. \ref{sec:scars}, are displayed in Fig. \ref{fig:single_qmbs_entropy}. Starting in a scar state leads to no entanglement growth, whereas the entanglement entropy for the chosen nonscar state quickly grows and saturates around the Page value. These results are consistent with Hamiltonian and Floquet models which obey the ETH and contain a vanishingly small fraction of scars \cite{Ser-21a}.

\begin{figure}
    \centering
    \includegraphics[width=0.9\linewidth]{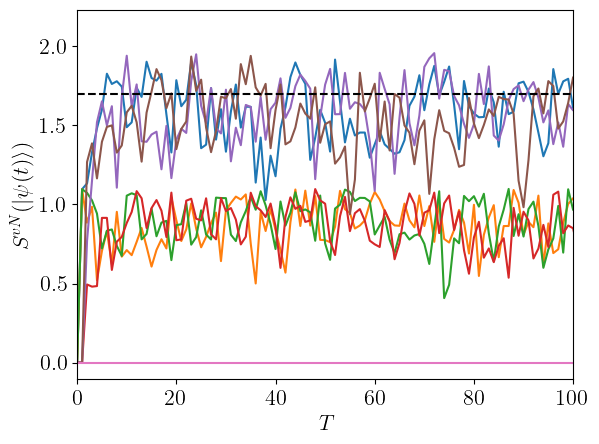}
    \caption{Growth of bipartite entanglement entropy for a set of chosen initial states, using the HSF model with $N = 6, \; d = 3$. For the frozen states (pink), $S^{vN}$ stays at $0$ for all times $t$. When starting in states belonging to the largest, $15$-dimensional subspace (blue, purple, brown), the entanglement entropy quickly achieves, and oscillates around the Page value (dashed black line). The states belonging to $7$-dimensional subspaces (red, green, orange), oscillate close around half of the Page value.}
    \label{fig:hsf_ee_growth}
\end{figure}

Finally, the effects of fragmentation can also be seen in the growth of bipartite entanglement entropy, displayed in Fig. \ref{fig:hsf_ee_growth}. If the system is initialized in one of the frozen states, then it remains in the frozen state for all times, and therefore the entanglement entropy stays at $0$. However, if the system is initialized in any state in the largest subspace, then it will quickly increase to the Page value and oscillate around it. 

\newpage

\begin{figure}
    \centering
    \includegraphics[width=\linewidth]{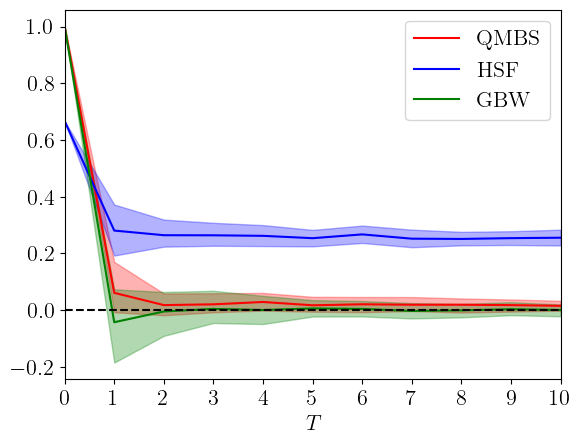}
    \caption{Comparison of the decay of the infinite temperature autocorrelator $\mathcal{A}$, as defined in Eq.~\eqref{eq:autocorrelator}, for the generic brickwork model (GBW), scarred model (QMBS), and the model with Hilbert space fragmentation (HSF). For all of these we use system size $N = 6$. The local Hilbert space dimension is $d = 2$ for GBW and QMBS models, while we use $d = 3$ for the HSF one. The local observable used to compute the autocorrelator for QMBS and GBW is $\hat{\sigma}^{z}_{N/2}$, and the one for HSF is $\hat{S}^{z}_{N/2}$, where $\hat{S}^z = \ket{1}\bra{1} - \ket{-1}\bra{-1}$. We observe that in the GBW the autocorrelator decays to almost exactly $0$, while for the QMBS it decays almost to $0$. Note that this small deviation is exponentially small in systems size and would vanish in the thermodynamic limit. In the case of HSF however, the autocorrelation function saturates to a finite bound of $1/3$, which is independent of system size, and indicates strong ergodicity breaking.}
    \label{fig:autocorrelator_comparison}
\end{figure}


\section{Decay of autocorrelation functions} \label{app:autocorrelators}

The second quantity which is relevant for the diagnosis of scars and fragmentation is the infinite temperature autocorrelator for a local operator $\hat{O}$, defined as:
\begin{equation} \label{eq:autocorrelator}
    \mathcal{A}(\hat{O}, t) = \text{Tr}[\frac{1}{D} \hat{\mathbb{I}} \hat{O}(t) \hat{O}(0)],
\end{equation}
where $\hat{O}(t) \equiv \hat{U}(t)^{\dagger} \hat{O} \hat{U}(t)$, and $\frac{1}{D} \hat{\mathbb{I}}$ represents the infinite temperature state. For a scarred model, this quantity should decay to almost $0$ with time as the scars represent a vanishing fraction of the Hilbert space, whereas a model exhibiting strong HSF will decay to a nonzero value, which is bounded from below by the Mazur bound \cite{Mou-22a}.

In Fig. \ref{fig:autocorrelator_comparison}, we see the decay of the autocorrelator for the HSF model presented in this section, contrasted with the models presented in Secs. \ref{sec:bw_chse} and \ref{sec:scars}. Note that, due to different dimensions of the local Hilbert spaces, we use the local observable $\hat{S}^z = \ket{1}\bra{1}  - \ket{-1}\bra{-1}$ for the former model, and the usual Pauli $Z$ operator for the latter two. The results are congruent with Hamiltonian models that display fragmentation. The autocorrelator quickly decays to a finite bound of $1/3$ and stays well above the long time values for the QMBS, which saturates at a very small finite value, and generic brickwork (GBW) models which saturate at exactly $0$.

\begin{figure} [b]
    \centering
    \includegraphics[width=0.8\linewidth]{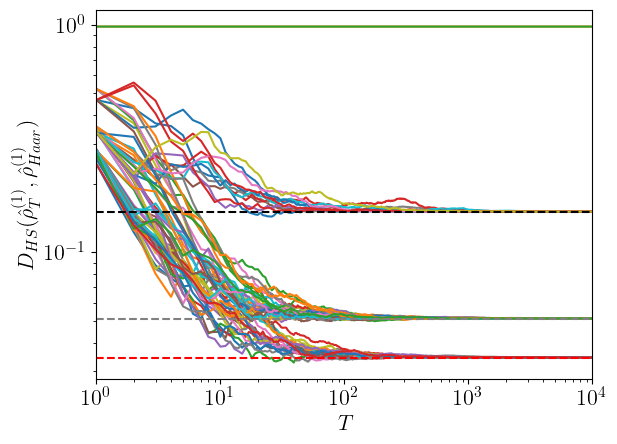}
    \includegraphics[width=0.8\linewidth]{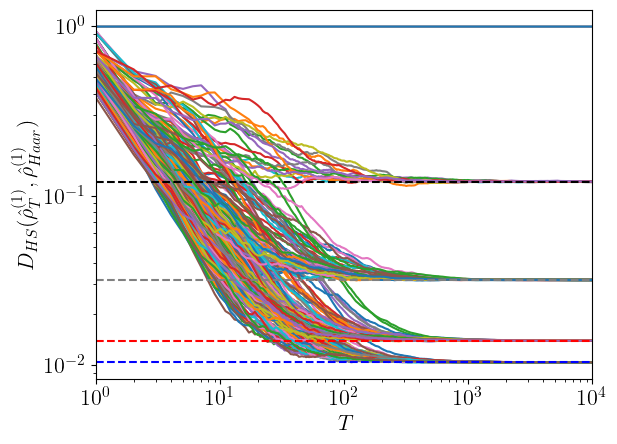}
    \caption{Hilbert-Schmidt distance between the first moments of the Haar ensemble, and the temporal ensemble generated by the Fibonacci layered pair-flip model, with $d = 2$, $N = 6$ (top) and $N = 8$ (bottom). We test all possible initial product states. The dashed lines represent the bounds imposed by different subspaces, and we observe convergence of each of the initial states to their respective bounds.}
    \label{fig:additional_symmetry_results}
\end{figure}

\section{Additional results for symmetries}\label{app:symmetries}

In the main text we considered the simple example of a $U(1)$ symmetry present in the pair-flip model, and we presented results for $N = 4$, for the two moments $k = 1, 2$. Here we present additional results for greater system sizes ($N = 6, 8$), but only computing the distance to the first Haar moment. In Fig. \ref{fig:additional_symmetry_results} we again observe that the distances of the different moments are converging to the bounds imposed by the subspace sizes.

\section{Additional Results for HSF}

\begin{figure} 
    \centering
    \includegraphics[width=0.8\linewidth]{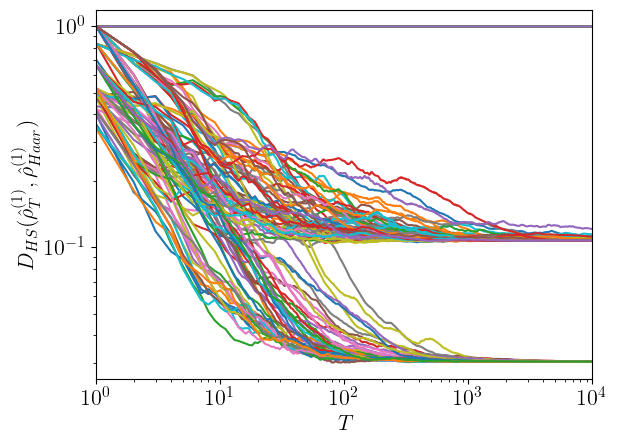}
    \includegraphics[width=0.8\linewidth]{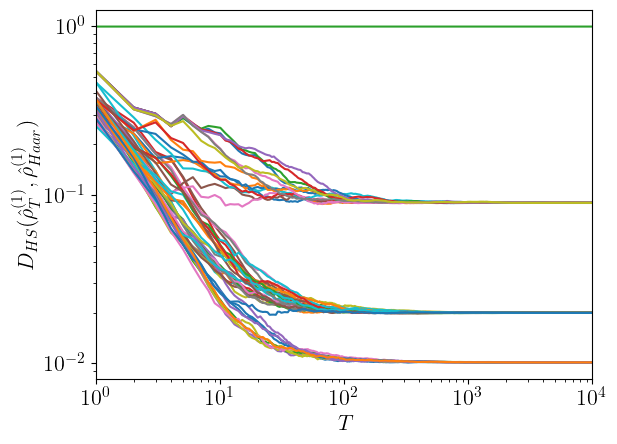}
    \caption{Hilbert-Schmidt distance between the first moments of the Haar ensemble, and the temporal ensemble generated by the Fibonacci layered pair-flip model, with $d = 3$, $N = 5$ (top) and $N = 6$ (bottom). We test half of all possible initial product states for $N = 5$ and $50$ initial product states for $N = 6$ to generate the temporal ensembles.}
    \label{fig:additional_hsf_results}
\end{figure}

In Sec. \ref{sec:fragmentation} of the main text we considered the pair-flip model with $d =3$ and $N = 4$, for the two moments $k = 1, 2$. Here we present additional results for $N = 5, 6$, with $k = 1$. In Fig. \ref{fig:additional_hsf_results} we again observe that the distances of the different moments are converging to the bounds imposed by the subspace sizes. 

\begin{figure*} 
    \centering
    \includegraphics[width=0.49\linewidth]{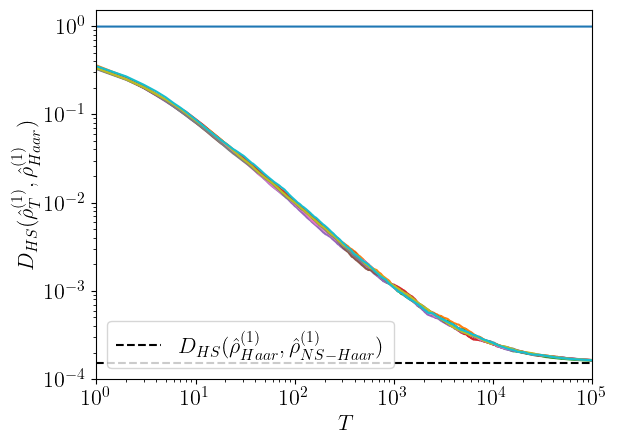}
    \includegraphics[width=0.49\linewidth]{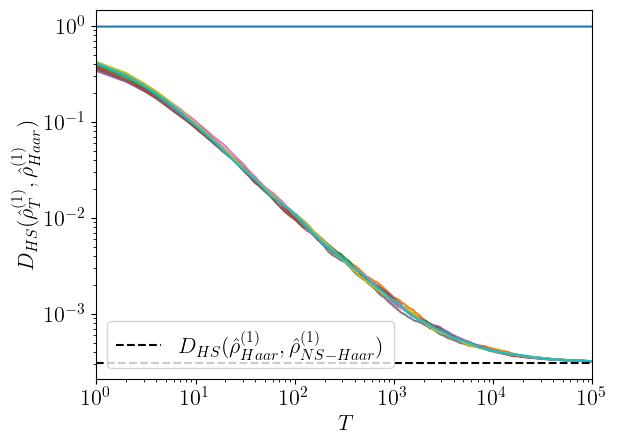}
    \includegraphics[width=0.49\linewidth]{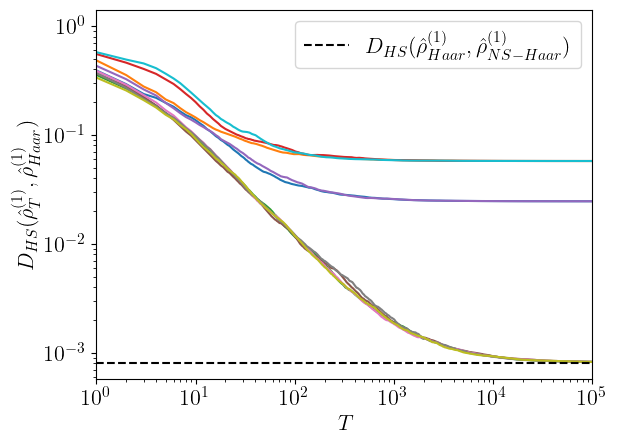}
    \includegraphics[width=0.49\linewidth]{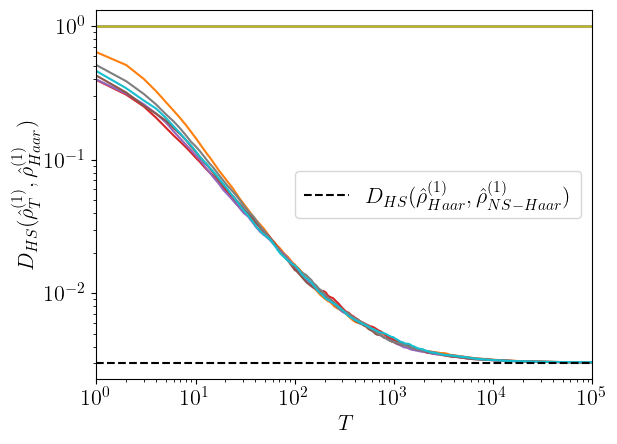}
    \caption{Results comparing the distances between the Haar ensemble and temporal ensemble, with $k = 1$, for the local unitaries of the form $\hat{U} = \text{exp}[i\hat{P}\hat{H}\hat{P}]$. We employ four different projectors $\hat{P}_1, \; \hat{P}_2, \; \hat{P}_{\text{lin}}$ and $\hat{P}_{\text{exp}}$, as defined in Eq. \eqref{eq:extra_projectors}. The results for the different projectors are displayed in that respective order, top to bottom. Note that increasing the number of embedded scars also leads to a decrease in the available space to explore for the other states. Therefore, a greater number of QMBS directly leads to lower bound on the distance, which can be observed from the results. Note that in the linear scar case, the QMBS states are not computational basis states, therefore they are affected by the actions of the local unitaries.}
    \label{fig:more_scars}
\end{figure*}

\section{Embedding Multiple Scars} \label{app:more_scars}

In the main part of our work we presented the simplest example of a single scar in a model with $N = 4$, $d = 2$. However, it is relatively straightforward to extend the presented results to other cases. For this purpose, we here consider a model with $N = 4$ qutrits ($d = 3$). We again perform simulations with the birckwork model, where the $2$-local unitaries have the form presented in Eq. \eqref{eq: u_scar}. We use four different projectors:

\begin{align} 
    \hat{P}_1 = \hat{\mathbb{I}} - \ket{00}\bra{00}, \hat{P}_2 = \hat{\mathbb{I}} - \ket{00}\bra{00} - \ket{11}\bra{11} \notag \\
    \hat{P}_{\text{exp}} = \hat{\mathbb{I}} - \ket{00}\bra{00} - \ket{11}\bra{11} - \ket{01}\bra{01} - \ket{10}\bra{10} \notag \\
    \hat{P}_{\text{lin}} = \hat{\mathbb{I}} - \ket{00}\bra{00} - \ket{11}\bra{11} - \ket{\Phi _{-}} \bra{\Phi _{-}} \notag \\
    \ket{\Phi _{-}} \equiv \ket{+-} - \ket{-+}, \: \ket{\pm} = \ket{0} \pm \ket{1}. \label{eq:extra_projectors}
\end{align}
Now, using $\hat{P}_1$ in the embedding procedure will result in a single scar, $\hat{P}_2$ in two scars, $\hat{P}_{\text{exp}}$ in $2^N$ scars and $\hat{P}_{\text{lin}}$ will give a linear number, $N + 1$, of scars. We perform numerical simulations for all four cases, for times up to $10^4$ and compare the first moments to the Haar ensemble. The results are presented in Fig. \ref{fig:more_scars}.

\end{document}